\documentclass[12pt]{article}

\usepackage{amssymb}
\usepackage{color}
\usepackage{amssymb,amsfonts,amsmath,graphicx,dsfont,cite,xfrac}
\usepackage{authblk}
\usepackage{subfigure}
\definecolor{mygray}{gray}{0.5}

\usepackage{cite}
\usepackage[colorlinks=true,linkcolor=blue,citecolor=red]{hyperref}

\newcommand{\be}{\begin{equation}}
\newcommand{\ee}{\end{equation}}
\newcommand{\bea}{\begin{eqnarray}}
\newcommand{\eea}{\end{eqnarray}}

\title{Nonclassical States for Non-Hermitian Hamiltonians with the Oscillator Spectrum}

\author[${1}$,${2}$]{K. Zelaya\thanks{zelayame@crm.umontreal.ca}}
\author[${3}$]{S. Dey\thanks{dey@iisermohali.ac.in}}
\author[${1}$,${4}$]{V. Hussin\thanks{hussin@dms.umontreal.ca}}
\author[${2}$]{O. Rosas-Ortiz\thanks{orosas@fis.cinvestav.mx}}

\affil[${1}$]{\footnotesize Centre de Recherches Math\'ematiques, Universit\'e de Montr\'eal, QC, H3C 3J7, Canada}

\affil[${2}$]{\footnotesize Physics Department, Cinvestav, AP 14-740, 07000
M\'exico City, Mexico}

\affil[${3}$]{\footnotesize Indian Institute of Science Education and Research  Mohali, Knowledge City, Sector 81, SAS Nagar (Mohali), PO Manauli, Punjab 140306, India }

\affil[${4}$]{\footnotesize D\'epartement de Math\'ematiques et de Statistique, Universit\'e de Montr\'eal, QC, H3C 3J7, Canada}

\date{}
\begin{document}

\maketitle

\begin{abstract}
We show that the standard techniques that are utilized to study the classical like properties of the pure states for Hermitian systems can be adjusted to investigate the classicality of pure states for non-Hermitian systems. The method is applied to the states of complex-valued potentials that are generated by Darboux transformations and can model both non-$PT$-symmetric and $PT$-symmetric oscillators exhibiting real spectra.
\end{abstract}


\section{Introduction}

In quantum mechanics one finds two important classes of states of the radiation field: Fock and coherent states. The former, introduced by Fock at the dawn of quantum theory, contain a precise number of photons and produce average fields equal to zero. With exception of the vacuum, the Wigner distribution \cite{Wig32} exhibits negative values in some regions of the  phase-space when it is evaluated with the Fock states. Thus, the properties of the quantum states of radiation fields that are occupied by a finite number of photons are far from the Maxwell theory. On the other hand, the coherent states were formally introduced in quantum optics by Glauber \cite{Gla07}, although the first antecedents can be traced back to the Schr\"odinger papers on quantization, see e.g. \cite{Ros19}. They produce average fields different from zero as well as nonnegative symmetrical Wigner distributions. Although the coherent states are constructed as superpositions of Fock states, it is remarkable that their properties are very close to the Maxwell theory. This is because such states satisfy the notion of full coherence introduced by Glauber, while the Fock states (for $n \neq 0$) lack second (and higher) order coherence, and thus they are nonclassical. The set of nonclassical states includes also squeezed states \cite{Wal83}, even and odd coherent states (also called Schr\"odinger cats) \cite{Dod74}, binomial states \cite{Sto85,Lee85}, photon-added coherent states \cite{Aga91}, etc. The difference between the classical and nonclassical properties of a given quantum state is strongly linked to the notion of entanglement \cite{Acz03}, which is a fundamental concept required in the development of quantum computation and quantum information \cite{Nie00}. Noticeably, while investigating the quantum properties of a beam splitter, it has been found that the ``entanglement of the output states is strongly related to the nonclassicality of the input fields''~\cite{Kim02}. That is, the nonclassicality of an arbitrary (input) state can be tested by detecting entanglement in the output of a quantum beam splitter~\cite{Kim02,Sch01}.

In the present work we study the nonclassical properties of the states of non-Hermitian Hamiltonians with real spectrum that are generated by Darboux transformations \cite{Ros15,Bla19}. In general, the systems associated with non-Hermitian Hamiltonians are subject of investigation in many branches of contemporary physics \cite{Bag16}. The applications include the study of unstable (decaying) systems \cite{Moy11}, light propagating in materials with complex refractive index \cite{Mos12}, $PT$-symmetric~\cite{Ben07,Fer98} and $PTC$-symmetric~\cite{Fak09} interactions, multi-photon transition processes \cite{Fai87}, diverse measurement techniques \cite{Sim17}, and coherent states \cite{Ros19,Dey18,Ros18}, among others. In the case of the Darboux deformed non-Hermitian systems, the eigenfunctions obey a series of oscillation theorems \cite{Jai17} that permit their study as if they were associated to Hermitian systems. Indeed, the representation space of such non-Hermitian systems can be equipped with a bi-orthogonal structure that provides complete sets of orthonormal states \cite{Ros15}. We shall focus on non-Hermitian oscillators since their bases of eigenfunctions can be used to construct optimized binomial states \cite{Zel16} and generalized coherent states \cite{Ros18}, which in turn give rise to a wide diversity of pure states as particular cases. One of the main results reported in this paper is to show that the techniques used to study classical like properties for Hermitian systems can be adjusted to investigate the classicality of the states of Darboux deformed non-Hermitian systems. 

The outline of the paper is as follows: In Section~\ref{secnonher} we revisit the main properties of the non-Hermitian oscillators we are interested in. The main differences between the conventional orthogonality and the bi-orthogonality are discussed in detail. In Section~\ref{secbi} we introduce the different bi-orthogonal superpositions of states that are to be analyzed, these include the optimized binomial states and diverse forms of generalized coherent states. Section~\ref{secnonclass} deals with the nonclassicality of the states defined in Section~\ref{secbi}. Final comments are given in Section~\ref{conclu}. We have added three appendices with relevant information that is used throughout the manuscript concerning operator algebras (Appendix~\ref{ApA}), and criteria of nonclassicality (Appendix~\ref{ApB}).

\section{Non-Hermitian oscillators}
\label{secnonher}

\begin{figure}[htb]
\centering
\subfigure[Non $PT$-symmetric]{\includegraphics[width=0.3\textwidth]{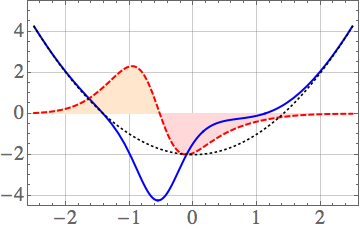}} 
\hspace{.5ex}
\subfigure[$PT$-symmetric]{\includegraphics[width=0.3\textwidth]{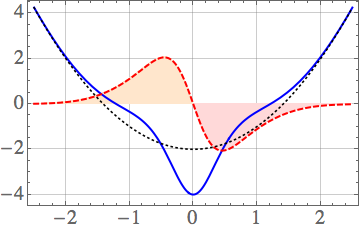}}
\hspace{.5ex}
\subfigure[Hermitian]{\includegraphics[width=0.3\textwidth]{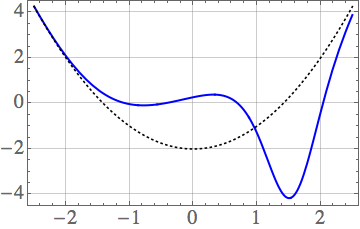}}

\caption{\footnotesize 
(Color online) The oscillators $V_{\lambda}(x)$ defined in (\ref{pot1}) can be categorized in three different classes: (a) non $PT$-symmetric (b) $PT$-symmetric and (c) Hermitian. In (a) and (b) the imaginary part (dashed-red) satisfies the condition of zero total area (\ref{zero}). The zones of gain and loss of probability are shadowed in orange and red respectively. In all the cases the curve in blue corresponds to the real part of $V_{\lambda}(x)$ and the dotted-black curve represents the shifted oscillator $V_{osc}(x)-2$, the latter is obtained from (\ref{pot1}) as a particular case.
}
\label{FigPot}
\end{figure}

The solution to the eigenvalue problem of the dimensionless (mathematical) oscillator $H_{osc}=-\frac{d^2}{dx^2} + x^2$ is given by the discrete eigenvalues $E_n^{osc}=2n+1$ and normalized eigenfunctions
\begin{equation}
\varphi_{n}(x)= \frac{e^{-x^{2}/2}}{\sqrt{2^{n}n!\sqrt{\pi}}} H_n(x), \quad n=0,1,\ldots,
\label{sol3}
\end{equation}
where $H_n(x)$ stands for the Hermite polynomials \cite{Olv10}. Using the linearly independent solutions of the eigenvalue equation $H_{osc}u=-u$ and a set of real numbers $\{ a,b, c, \lambda\}$ fulfilling $4ac-b^2 = 4 \lambda^2$, it can be shown that 
\begin{equation}
\alpha(x) =  e^{x^2/2}  \left[ \frac{a \pi}{4} \mbox{Erf}^2(x) +  \frac{b \sqrt{\pi}}{ 2} \mbox{Erf}(x) + c \right]^{1/2}
\label{pot2}
\end{equation}
is a real-valued function which is free of zeros in $\mathbb R$, and defines the dimensionless potential
\begin{equation}
V_{\lambda}(x) =x^2 -2- 4 \frac{d}{dx}\left[ \frac{  a \sqrt{\pi} \, \mbox{Erf}(x) +b -  i  \frac{\lambda}{2} }{\alpha^{2}(x)} \right]
\label{pot1}
\end{equation}
as a Darboux transformation of $V_{osc}(x)=x^2$ \cite{Ros15}. Here Erf$(x)$ defines the error function. If $\lambda \neq 0$, the Hamiltonian $H_{\lambda} = -\frac{d^2}{dx^2} + V_{\lambda}(x)$ is not self-adjoint since  $V_{\lambda \neq 0} (x)$ is  a complex-valued function. Nevertheless, it may be shown that the imaginary part of such potential is continuous in $\mathbb R$ and satisfies the condition of {\em zero total area} \cite{Jai17}:
\begin{equation}
\int_{\mathbb R} \mbox{Im} V_{\lambda} (x) dx = - \left. \frac{2\lambda}{\alpha^2(x)} \right\vert_{-\infty}^{+\infty}=0.
\label{zero}
\end{equation}
The condition (\ref{zero}) implies a balanced interplay between gain and loss of probability that does not depend on any other symmetry of either $\mbox{Im} V_{\lambda}(x)$ or $\mbox{Re} V_{\lambda}(x)$. For instance, besides fulfilling (\ref{zero}), the profile of the real and imaginary parts of the oscillator depicted in Fig.~\ref{FigPot}(a) is not particularly special (this has been obtained from (\ref{pot1}) with $a=5$, $b=8$, $c=5$ and $\lambda= 3$). Looking for concrete symmetries we can take $b=0$ to get even and odd forms of $\mbox{Im} V_{\lambda}(x)$ and $\mbox{Re} V_{\lambda}(x)$ respectively. The result is a non-Hermitian oscillator which is invariant under $PT$-transformations \cite{Ben07}. This case is illustrated in Fig.~\ref{FigPot}(b) for $a=c=\lambda=1$, and $b=0$. On the other hand, for $b=\pm 2\sqrt{ac}$, with $a$ and $c$ positive numbers, the parameter $\lambda$ is equal to zero and the expression (\ref{pot1}) decouples into the pair of real-valued functions
\begin{equation}
V_{\lambda=0}^{\pm} (x; \gamma) = x^2 -2 -2\frac{d}{dx} \left(\frac{e^{-x^2}}{ \int^{x} e^{-y^2}dy \pm \gamma} \right), \quad \gamma= \sqrt{\frac{c}{a}},
\label{potlim}
\end{equation}
where the condition $ \gamma > \sqrt{\pi}/2$ must be satisfied to avoid singularities. Fig.~\ref{FigPot}(c) shows $V_{\lambda=0}^{-} (x; \gamma)$ with $a=1$ and $c=0.88$. The concomitant $V_{\lambda=0}^{+} (x; \gamma)$ corresponds to the specular reflection of $V_{\lambda=0}^{-} (x; \gamma)$. The Hermitian oscillators (\ref{potlim}) were first found by Abraham and Moses through the systematic use of the Gelfand-Levitan equation \cite{Abr80}, and then recovered by Mielnik as an application of his generalized factorization method \cite{Mie84}. 

Additionally, the expressions (\ref{potlim}) represent a family of systems that converge to the conventional oscillator in the limit $\vert \gamma \vert \rightarrow \infty$. Indeed, it may be verified that
\begin{equation}
\lim_{\vert \gamma \vert \rightarrow \infty} V_{\lambda=0} (x; \gamma) \rightarrow x^2 -2.
\label{osc1}
\end{equation}
The oscillator $V_{osc}(x)-2$ is thus a member of the family of complex-valued potentials $V_{\lambda}(x)$ introduced in (\ref{pot1}). This is depicted in the plots of Fig.~\ref{FigPot} as a reference.

$\bullet$ {\bf Fundamental solutions.} Another remarkable profile of the complex-valued oscillators (\ref{pot1}) is that the functions
\begin{equation}
\psi_{n+1}(x) = \left[ \frac{d}{dx} -\frac{\alpha'(x)}{\alpha(x)}+i\frac{\lambda}{\alpha^{2}(x)} \right]\varphi_{n}(x),  \quad  n=0,1,\ldots, 
\label{psia}
\end{equation}
are normalizable solutions of the related eigenvalue equation $H_{\lambda} \psi_{n+1}(x) = E_{n+1} \psi_{n+1}(x)$ for $E_{n+1}=2n+1$. The additional normalizable function
\begin{equation}
\psi_{0}(x) =\frac{1}{\alpha(x)} \exp \left[ i \arctan \left( \frac{a \sqrt{\pi} \, \mbox{Erf}(x) +b}{2 \lambda}
 \right)
\right]
\label{psib}
\end{equation}
also solves the eigenvalue equation defined by $H_{\lambda}$, but it is not derivable from the set $\varphi_n(x)$ and belongs to the energy $E_0=-1$. The spectrum of the potential $V_{\lambda}(x)$ is therefore composed of the equidistant energies $E_n=2n-1$, $n \geq 0$. 

$\bullet$ {\bf Bi-orthogonality.} Although the functions $\psi_n(x)$ are normalizable, they form a peculiar set since $\psi_0(x)$ is orthogonal  to all the $\psi_{n+1}(x)$ while the latter are not mutually orthogonal. As a result, in contrast with the Hermitian case, the norm of any superposition of states $\psi_{n+1}(x)$ depends not only on the modulus of the related coefficients but also on the phase shift between them. For instance, the norm of 
\begin{equation}
\Psi(x)= \frac{\theta}{\vert \vert \psi_{n+1} \vert \vert} \, \psi_{n+1}(x) + \frac{\mu}{\vert \vert \,\psi_{m+1} \vert \vert} \psi_{m+1}(x), \quad \theta, \mu \in \mathbb C,
\label{phin}
\end{equation}
satisfies
\begin{equation}
\vert\vert \Psi \vert\vert^2 = \vert \theta \vert^2 + \vert \mu \vert^2 + 2 \vert \theta \mu \vert  \,\xi_{n+1, m+1}  \cos(\delta + \gamma_{n+1, m+1}), \quad \delta = \mbox{arg}(\theta^* \mu),
\label{cuatro}
\end{equation}
where $z^*$ is the complex-conjugate of $z\in \mathbb C$. The numbers $\xi_{n+1,m+1}$ and $\gamma_{n+1,m+1}$ are respectively the modulus and argument of the product between $ \psi_{n+1}(x)$ and $\psi_{m+1}(x)$. Thus, depending on the phase-shift $\delta$, the non-orthogonality of the set $\psi_{n+1}(x)$ produces the oscillations of $\vert \vert \Psi \vert \vert$. The complexity of such a dependence increases with the number of elements in the superposition. 

One can face the above difficulties by introducing a bi-orthogonal system (see relevant information in  Ref.~\cite{Ros18}), formed by the eigenfunctions $\psi_n(x)$ of $H_{\lambda}$ and those of its Hermitian-conjugate $H_{\lambda}^{\dagger} \equiv \overline H_{\lambda}$, written as $\overline \psi_m(x)$. The main point is that the bi-product $\left( \overline \psi_m, \psi_n \right)= \int_{\mathbb R}\overline \psi_m^{\, *} (x) \psi_n (x) dx$ is equal to zero if $n\neq m$, and serves to define the bi-norm $\vert \vert \psi_n \vert \vert_B = \vert \vert \overline \psi_n \vert \vert_B$ if $n=m$. 

\begin{figure}[htb]
\centering
\subfigure[$\psi_0(x)$]{\includegraphics[width=0.2\textwidth]{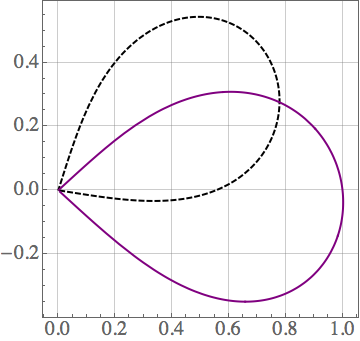}} 
\hskip2ex
\subfigure[$\psi_1(x)$]{\includegraphics[width=0.19\textwidth]{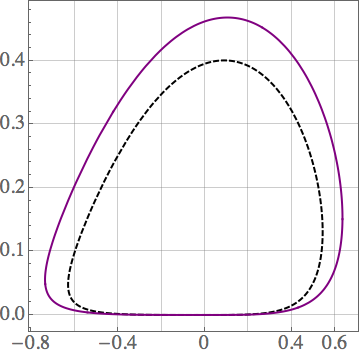}}
\hskip2ex
\subfigure[$\psi_2(x)$]{\includegraphics[width=0.21\textwidth]{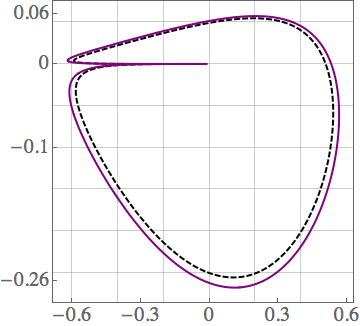}}
\vskip1ex
\subfigure[$\vert \psi_0(x) \vert^2$]{\includegraphics[width=0.2\textwidth]{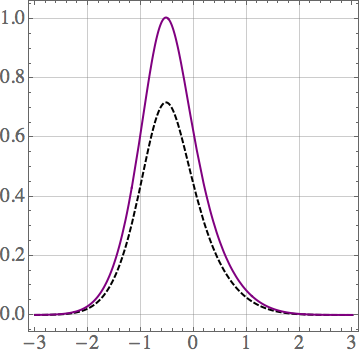}} 
\hskip2ex
\subfigure[$\vert \psi_1(x) \vert^2$]{\includegraphics[width=0.2\textwidth]{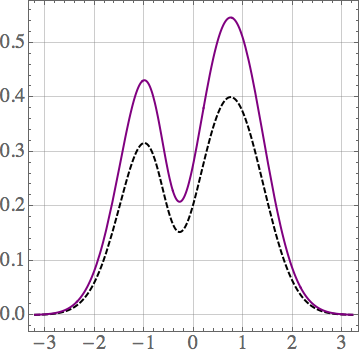}}
\hskip2ex
\subfigure[$\vert \psi_2(x) \vert^2$]{\includegraphics[width=0.2\textwidth]{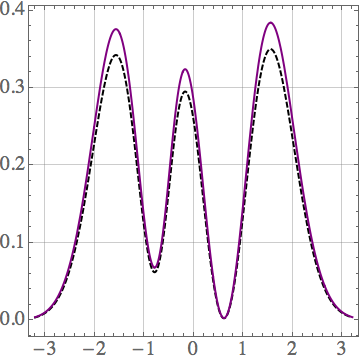}}

\caption{\footnotesize 
(Color online) Bi-normalization (continuous-purple) against conventional normalization (dotted-black) for the first three eigenfunctions $\psi_n(x)$ --upper row-- and probability densities $\vert \psi_n(x) \vert^2$ --lower row-- of the complex-valued oscillator (\ref{pot1}) shown in Fig.~\ref{FigPot}(a). The eigenfunctions are depicted in the Argand-Wessel representation with $\mbox{Re} \, \psi_n (x)$ and $\mbox{Im} \, \psi_n (x)$ in the horizontal and vertical axes, respectively. They have been constructed from (\ref{psia}) and (\ref{psib}) with $a=5$, $b=8$, $c=5$ and $\lambda=3$.
}
\label{Figdensity}
\end{figure}

Therefore, besides the conventional normalization $\psi_n(x)/\vert \vert \psi_n \vert \vert$, we have at hand the bi-normalization $\psi_n(x)/\vert \vert \psi_n \vert \vert_B$, with $\vert \vert \psi_{n+1} \vert\vert_B = \sqrt{2(n+1)}$ for $n\geq 0$. The bi-norm $\vert \vert \psi_0 \vert \vert_B$  of the ground state depends on the set $\{ a,b,c,\lambda\}$  \cite{Ros18}. The real and imaginary parts of $\psi_n (x)$, as well as the probability density $\vert \psi_n(x) \vert^2$, behave qualitatively equal in both normalizations but their bi-normalized values are usually larger than those obtained with the conventional normalization. Such a difference is reduced as $n$ increases. This property is illustrated in Fig.~\ref{Figdensity} for the first three eigenfunctions $\psi_n(x)$, and the corresponding probability densities $\vert \psi_n(x) \vert^2$, of the complex-valued oscillator depicted in Fig.~\ref{FigPot}(a).

\begin{figure}[htb]
\centering
\subfigure[$\delta=0.55$]{\includegraphics[width=0.2\textwidth]{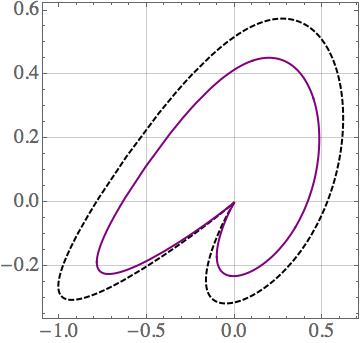}} 
\hskip2ex
\subfigure[$\delta=1.6$]{\includegraphics[width=0.2\textwidth]{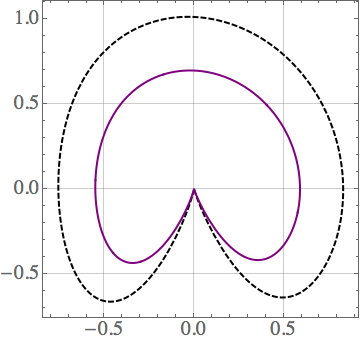}}
\hskip2ex
\subfigure[$\delta=2.28$]{\includegraphics[width=0.2\textwidth]{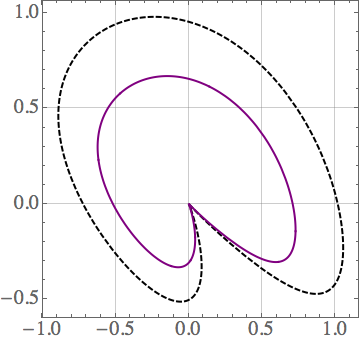}}
\vskip1ex
\subfigure[$\delta=0.55$]{\includegraphics[width=0.2\textwidth]{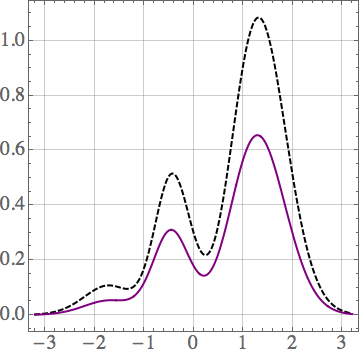}} 
\hskip2ex
\subfigure[$\delta=1.6$]{\includegraphics[width=0.2\textwidth]{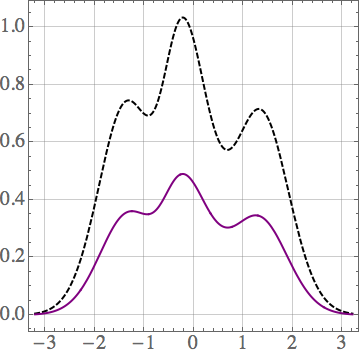}}
\hskip2ex
\subfigure[$\delta=2.28$]{\includegraphics[width=0.2\textwidth]{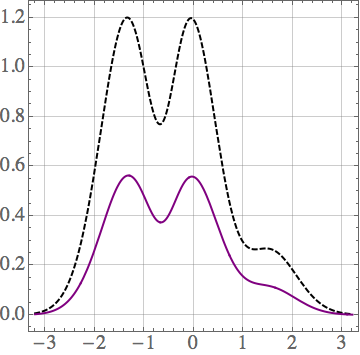}}

\caption{\footnotesize 
(Color online) The upper row shows the Argand-Wessel diagrams of the superpositions $\Psi(x)$ and $\Psi_B(x)$, Eqs. (\ref{phin}) and (\ref{phib}) respectively, for $n=0$, $m=1$, $\vert \theta \vert = \vert \mu \vert = 1/\sqrt 2$, and the indicated values of the phase-shift $\delta$. The corresponding probability densities are shown in the lower row. The code of colors and plot-style is the same as in Fig~\ref{Figdensity}.
}
\label{Figsuper}
\end{figure}

The bi-orthogonal approach avoids the interference produced by the non-orthogonality. For instance, if the sates in (\ref{phin}) are substituted by their bi-normalized versions, we obtain the bi-orthogonal superposition 
\begin{equation}
\Psi_B(x)= \frac{\theta}{\sqrt{2(n+1)} } \, \psi_{n+1}(x) + \frac{\mu}{\sqrt{2(m+1)} } \,  \psi_{m+1}(x).
\label{phib}
\end{equation} 
The bi-norm of the latter state does not depend on the phase-shift 
\begin{equation}
\vert\vert \Psi \vert\vert^2_B = \int_{\mathbb R} \overline \Psi_B^{\, *}(x) \Psi_B(x) dx=
\vert \theta \vert^2 + \vert \mu \vert^2,
\label{biprod}
\end{equation}
so that $\Psi_B(x)$ is uniquely bi-normalized to 1 by the constant $(\vert \theta \vert^2 + \vert \mu \vert^2)^{-1}$. An additional property of the bi-orthogonal superpositions is that the values of their real and imaginary parts, as well as the values of the corresponding probability densities, are shorter than those obtained from the conventional approach. We show this property in Fig.~\ref{Figsuper} for the superpositions (\ref{phin}) and (\ref{phib}) with $n=0$, $m=1$, $\vert \theta \vert = \vert \mu \vert = 1/\sqrt 2$, and three different values of the phase-shift $\delta$. Formally, one may say that $\Psi(x)$ and $\Psi_B(x)$ represent two different superpositions of the states $\psi_{n+1}(x)$ and $\psi_{m+1}(x)$. In the sequel we shall take full advantage of the mathematical simplifications offered by the bi-orthogonal approach.

$\bullet$ {\bf Operator algebras.} It can be shown  that there exist at least two different algebras of operators associated with the eigenstates of the complex-valued oscillator $V_{\lambda}(x)$ \cite{Zel19}. They are generated by two different pairs of ladder operators (see Appendix~\ref{ApA} for details). The first pair, ${\cal A}$ and ${\cal A}^+$, together with the Hamiltonian $H_{\lambda}$, satisfy the {\em quadratic polynomial}  (Heisenberg) algebra
\begin{equation}
[{\cal A}, {\cal A}^{+}]= 2 \left(3 H_{\lambda} +1 \right) \left( H_{\lambda} +1 \right), \quad [H_{\lambda}, {\cal A} ]=-2 {\cal A}, \quad [H_{\lambda}, {\cal A}^+ ]= 2 {\cal A}^+.
\label{nat2}
\end{equation}
The second pair of ladder operators, denoted by ${\cal C}_w$ and ${\cal C}_w^{+}$, together with the Hamiltonian $H_{\lambda}$, and an additional operator $I_w$, satisfy the {\em distorted} (Heisenberg) algebra
\begin{equation}
[{\cal C}_w, {\cal C}_w^{+} ] = I_w, \quad [H_{\lambda}, {\cal C}_w ]= -2 {\cal C}_w, \quad [H_{\lambda}, {\cal C}_w^+ ]= 2 {\cal C}_w^+,
\label{dist1}
\end{equation}
where $w$ is a non-negative parameter that defines the `distortion' suffered by the oscillator algebra when 
one substitutes the operators ${\cal C}_w$, ${\cal C}_w^{+}$, and $H_{\lambda}$ for the conventional boson operators, $\hat a$, $\hat a^{\dagger}$, and $\hat n$, respectively. The algebras (\ref{nat2}) and (\ref{dist1}) will serve to the analysis of classicality of the states associated with the complex-valued oscillators (\ref{pot1}).

\section{Bi-orthogonal superpositions}
\label{secbi}

We shall study the properties of diverse superpositions of states $\psi_n(x)$ by using 
\begin{enumerate}
\item[(i)]  Bi-normalization to obtain regular probability densities. 
\item[(ii)] Bi-orthogonality to avoid the interference associated with non-orthogonality.
\end{enumerate}
The space of states of $H_{\lambda}$, denoted ${\cal H}_{\lambda}$, consists of all the bi-orthogonal superpositions 
\begin{equation}
\phi(x)= \sum_{k=0}^{\infty} c_k \psi_k(x), \quad c_k= \left(\overline \psi_k, \phi \right) \in \mathbb C,
\label{super1}
\end{equation}
such that
\begin{equation}
\vert \vert \phi \vert \vert_B^2=(\overline \phi, \phi )=  \sum_{k=0}^{\infty} \vert c_k \vert^2 <\infty.
\label{norm}
\end{equation}
In turn, $\overline{\cal H}_{\lambda}$ denotes the space spanned by the states $\overline \psi_n(x)$. The series decomposition of $\overline \phi (x)$, which is concomitant to $\phi(x)$ in (\ref{norm}), is obtained from (\ref{super1}) by  changing $\psi_k(x) \rightarrow \overline \psi_k(x)$. To simplify the notation, hereafter we use no label to denote the bi-orthogonal properties of the system $\{ \overline \psi_n(x), \psi_m (x) \}$.

The approach we are going to use identifies a threefold partnership between $H_{\lambda}$, $\overline H_{\lambda}$, and the conventional oscillator $H_{osc}$, where the eigenvalues play a main role. That is, the triad $\{ \overline H_{\lambda}, H_{osc}, H_{\lambda} \}$ represents a system for which the energies $E_{n+1}=E^{osc}_n =2n+1$ are threefold degenerate while the energy $E_0=-1$ is only twofold degenerate. In this sense, our model is a generalization of the supersymmetric approach \cite{Mie04} since $H_{\lambda}$ and $\overline H_{\lambda}$ are treated as two different faces of the same system, which can be studied in much the same way as in the Hermitian approaches \cite{Ros18}.

We are interested in two different classes of superpositions. The {\em optimized superpositions} are such that their nonclassicality can be manipulated by tuning the number of elements as well as the state of lowest energy in the sum. The {\em generalized coherent states} are constructed in terms of the algebras underlying the complex-valued oscillators (\ref{pot1}) and form an over-complete set in ${\cal H}_{\lambda}$. Next, we analyze these classes in detail.

\subsection{Optimized binomial states}

Let us consider a superposition of $N+1$ adjacent states 
\begin{equation}
\phi_b(x)= \sum_{k=0}^N c_k^b \psi_{k+r}(x),
\label{finite}
\end{equation}
where the non-negative integer $r \geq 0$ determines the state $\psi_r (x)$ of the lowest energy that is included in the packet, and the coefficients $c_k^b$ give rise to the binomial distribution 
\begin{equation}
\vert c_k^b \vert^2= \left( 
\begin{array}{c}
N\\ k
\end{array}
\right) p^{k}(1-p)^{N-k}, \quad 0\leq p \leq 1, \quad k\leq N.
\label{finite2}
\end{equation} 
That is, the sum (\ref{finite}) is bi-normalized and includes only the eigenstates belonging to the energies $E_r, E_{r+1}, \ldots, E_{r+N}$. The distribution (\ref{finite2}) means that the probability of finding the system in the state $\psi_{k+r} (x)$ is weighted by the probability $p$ of having success $k$ times in $n$ trials. The superpositions (\ref{finite}) are called {\em optimized binomial} states \cite{Zel16} since (for $r=0$) they converge to the conventional binomial superposition at the oscillator limit.

To investigate the properties of $\phi_b(x)$ let us take $p \approx 0$. The modulus $\vert c_0^b \vert$ of the first coefficient in the expansion is close to 1 and the other coefficients  are almost zero. Then, the state $\psi_r (r)$ is very influential in the behavior of the entire packet since it is weighted by $c_0^b$ in the superposition. The situation changes if $p \approx 1$  since in this case the modulus $\vert c_N^b \vert$ of the last coefficient in the sum is close to 1, while the moduli of the other coefficients are almost zero. In such case, $\psi_{N+r} (x)$ is the state of major influence in the packet. 

Another important feature of the optimized binomial states is that their mean energy ${\cal E}_b$ depends on the number $r$ which labels the lowest of the involved energies  
\begin{equation}
{\cal E}_b= \langle H_{\lambda}\rangle_{b} =2(Np+ r) -1.
\label{enbin}
\end{equation}
Considering $0\leq p \leq 1$, we have  
\begin{equation}
2r-1 \leq {\cal E}_b \leq 2(N + r)-1.
\label{bound}
\end{equation}
That is, ${\cal E}_b$ is bounded from below by $r$, and from above by $N$ and $r$. Fixing ${\cal E}_b$ and $r$ in (\ref{enbin}) we can write
\begin{equation}
Np = \tfrac12 \left( {\cal E}_b +1-2r \right).
\label{bin}
\end{equation}
Then, from (\ref{finite2}) we arrive at the Poisson distribution:
\begin{equation}
\lim_{N \rightarrow +\infty} \vert c_k^b \vert^2 = \frac{e^{-\frac{{\cal E}_b +1-2r}{2}}}{k!}
\left( \frac{{\cal E}_b +1-2r}{2} \right)^k.
\end{equation}
In this form,  as a limit version of the optimized binomial state $\phi_b (x)$, we introduce the {\em optimized Poisson} state
\begin{equation}
\phi_{P} (x) = e^{-\frac{\vert z \vert^2}{2}} \sum_{k=0}^{\infty} \frac{z^k}{\sqrt{k!}}  \psi_{k+r} (x), \qquad \vert z \vert^2= \frac{{\cal E}_b +1-2r}{2}. 
\label{poisson}
\end{equation}
In the quantum oscillator limit (\ref{osc1}), the superposition (\ref{poisson}) acquires the form 
\begin{equation}
\phi_P^{osc} (x) = \left. \lim_{\vert \gamma \vert \rightarrow +\infty}  \phi_P(x) \right \vert_{\lambda=0}=  \sum_{k=0}^{+\infty}  c_k^P \varphi_{k+r} (x), \quad c_k^P= e^{-\frac{\vert z \vert^2}{2}}  \frac{z^k}{\sqrt{k!}},
\label{opclas2}
\end{equation}
which coincides with the conventional coherent state for $r=0$, see Eq.~(\ref{glauber}) of Appendix~\ref{ApB}. 

\subsection{Generalized coherent states}
\label{seccs}

The ground state $\psi_0 (x)$ is annihilated by both pairs of ladder operators ${\cal A}$, ${\cal A}^+$, and ${\cal C}_w$, ${\cal C}_w^+$, see Appendix~\ref{ApA}. In any case we can construct coherent states as eigenstates of the respective annihilator operator. We write
\begin{equation}
\phi^{(\gamma)} (x) = \sum_{k=0}^{+ \infty} c_k^{(\gamma)}  \psi_{k+1} (x), \quad \gamma = {\cal N}, w, {\cal N}_d, w_{d}.
\label{super2}
\end{equation}
The super-index $\gamma$ stands for natural (${\cal N}$), distorted ($w$), natural displaced (${\cal N}_d$), and distorted displaced ($w_{d}$). Such nomenclature refers to the form in which the coefficients $c_k^{(\gamma)}$ have been selected, as it is explained below. For simplicity, we use Dirac notation.

If the superposition (\ref{super2}) is an eigenvector of the natural annihilation operator ${\cal A}$ with complex eigenvalue $z$, then it is a {\em natural coherent state} \cite{Ros18}. We write\footnote{A factor $\sfrac12$ of the complex eigenvalue reported in \cite{Ros18} has been absorbed in $z$.}
\begin{equation}
\vert z \rangle = \vert \phi^{({\cal N})} \rangle, \quad c_k ^{({\cal N})}= \frac{1}{\sqrt{ {}_0F_2(1,2,\vert z \vert^2)}} \, \frac{z^k}{k! \sqrt{(k+1)!}}.
\label{natural}
\end{equation}
Notice that also the ground state $\vert \psi_0 \rangle$ is eigenvector of ${\cal A}$, but its complex eigenvalue is equal to zero. Moreover, as $\vert z=0 \rangle = \vert \psi_1 \rangle$, we see that the eigenvalue $z=0$ is twice degenerate. The straightforward calculation shows that the vectors $\vert \psi_0 \rangle$ and $\vert z \rangle$ minimize the uncertainty relation (\ref{quadnat2}). In this sense $\vert \psi_0 \rangle$ and $\vert z \rangle$ represent two different types of minimal uncertainty states. 

Similarly, the {\em distorted coherent states} \cite{Ros18} are defined as eigenvectors of the distorted annihilation operator ${\cal C}_w$ with complex eigenvalue $z$. In this case we write
\begin{equation}
\vert z,w\rangle = \vert \phi^{(w)} \rangle, \quad c_k^{(w)} =\frac{1}{\sqrt{ {}_1F_1(1,w,\vert z \vert^2)}} \, \frac{z^k}{\sqrt{(w)_k}},
\label{distorted}
\end{equation}
with $(w)_k = w(w-1) \cdots (w-k+1)$ being the Pochhammer symbol \cite{Olv10}. The eigenvalue $z=0$ is also twice degenerate because $\vert \psi_0 \rangle$ and $\vert z=0, w \rangle=\vert \psi_1 \rangle$ are annihilated by ${\cal C}_w$. As $\vert \psi_0 \rangle$ and $\vert z, w \rangle$ minimize the uncertainty relation (\ref{distquad2}), they represent two different types of minimal uncertainty states.

Another interesting superposition (\ref{super2}) is obtained by demanding that the vector $\vert \phi^{(\gamma)} \rangle$ be the displaced version of a fiducial state. The states $\vert \psi_0 \rangle$ and $\vert \psi_1 \rangle$ are invariant under the action of the operator exponentiations $e^{-z^* {\cal A}}$ and $e^{-z^* {\cal C}_w}$, because they are annihilated by both operators ${\cal A}$ and ${\cal C}_w$. The latter means that $\vert \psi_0 \rangle$ and $\vert \psi_1 \rangle$ can be used as fiducial states. Considering the operators
\begin{equation}
D (z) = e^{z {\cal A}^+} e^{-z^* {\cal A} }, \qquad D_w(z) =e^{z {\cal C}^{+}_w} e^{-z^* {\cal C}_w},
\end{equation}
one has the `displaced' states $\vert \phi^{({\cal N}_{d})} \rangle = D (z) \vert \psi_1 \rangle$ and $\vert \phi^{(w_{d})} \rangle = D_w(z) \vert \psi_1 \rangle$. Remark that the ground state $\vert \psi_0 \rangle$ is invariant under the action of $D(z)$ and $D_w(z)$, so it is also a displaced state.

In the rest of this work we shall omit the description of the states $\vert \phi^{({\cal N}_{d})} \rangle$ since their properties are qualitatively similar to those of the (distorted) {\em displaced coherent states}:
\begin{equation}
\vert \psi_0 \rangle, \qquad \vert z,w\rangle_d = \vert \phi^{(w_d)} \rangle, \quad c_k^{(w_d)} =\frac{1}{\sqrt{ {}_1F_1(w,1,\vert z \vert^2)}} \, \frac{z^k \sqrt{(w)_k}}{k!}.
\label{displaced}
\end{equation}
In contrast to the previous superpositions, the states $\vert z, w \rangle_d$ do not minimize the uncertainty relation (\ref{distquad2}) for arbitrary values of $z$, but only for $z=0$. The latter means that $\vert \psi_0 \rangle$ and $\vert \psi_1 \rangle$ are the only displaced coherent states of minimal uncertainty.

\section{Nonclassical states for non-Hermitian oscillators}
\label{secnonclass}

The $P$-representation introduced by Glauber \cite{Gla63} and Sudarshan \cite{Sud63} defines a limit for the classical description of radiation fields. If the $P$-function is not accurately interpretable as a probability distribution then the field ``will have no classical analog'' \cite{Gla07}. This notion of {\em nonclassicality} applies immediately to the Fock states $\vert n \rangle$ for $n \neq 0$ because their $P$-functions are as singular as the derivatives of the delta function $\delta^{(2)}(z) = \delta(\mbox{Re}(z))\delta(\mbox{Im}(z))$, and are negative in some regions of the complex $z$-plane. The latter means that the light beams represented by any of the Fock states $\vert n \neq 0 \rangle$ cannot be described in terms of the electrodynamics introduced by Maxwell. In turn, the Glauber states \cite{Gla07}, usually denoted $\vert \alpha \rangle$, are {\em classical} because their $P$-function is precisely $\delta^{(2)}(z -\alpha)$. The vacuum $\vert 0 \rangle$ is also classical as it is a coherent state with complex eigenvalue equal to zero. In general, the states  for which the quadrature variances obey the inequalities 
\begin{equation}
(\Delta \hat x)^2 \geq \tfrac12, \quad (\Delta \hat p)^2 \geq \tfrac12,
\label{ineq}
\end{equation}
admit a non-negative $P$-function \cite{Gla07}. Here, $(\Delta A)^2= \langle A^2\rangle -\langle A \rangle^ 2$ stands for the variance of the operator $A$. If either of the inequalities (\ref{ineq}) is not satisfied, the $P$-function is ill-defined and the state is called {\em squeezed} \cite{Hol79,Cav81,Wal83}. In these cases it is better to represent the state using the Wigner distribution \cite{Wig32}, which is regular and always exists. Nonclassical states have Wigner functions which either are negative in some regions of the phase-space or are squeezed in one of the variables of the phase-space.

In modern days, the nonclassicality of a state can be tested by another excellent method, which follows from the use of the quantum beam splitter. It creates entangled states at its output ports while at least one of its inputs are fed with a nonclassical state. The device is used frequently both in theory and experiment, since it not only tests the nonclassicality of the state, but also quantifies the degree of nonclassicality in an efficient way \cite{Sch01,Kim02}. There are some well-known nonclassical states in the literature, which include squeezed states \cite{Wal83}, even and odd coherent states (also called Schr\"odinger cats) \cite{Dod74}, binomial states \cite{Sto85,Lee85}, photon-added coherent states \cite{Aga91}, etc. Apart from the nonclassical states that arise from the harmonic oscillator, there exist many other such states emerging from different generalizations of coherent states; see, for instance \cite{Dod03}. Other interesting nonclassical states have also been shown to originate from some mathematical frameworks, in particular, from the noncommutative systems reported in \cite{San15,San15b,San16}. The striking feature of such systems is that they give rise to well-defined nonclassical states although the corresponding models are non-Hermitian. 

Quite recently it has been shown that the appropriate generalizations of the oscillator algebra permit the construction of nonlinear coherent states that satisfy a closure relation which is expressed uniquely in terms of the Meijer $G$-function \cite{Zel17}. This property automatically defines the delta distribution as the corresponding $P$-representation. However, in the same work, it is also shown that does not exist a classical analog for such states since they lack second-order coherence and exhibit antibunching. Thus, although their $P$-representation is a delta function, the nonlinear coherent states studied in \cite{Zel17} are not classical since they are not fully coherent in the sense established by Glauber \cite{Gla07}. Similar results have been obtained for the para-Bose oscillator \cite{Moj18a,Moj18b}. The algebras (\ref{nat2}) and (\ref{dist1}) that we have used to construct the coherent states of Section~\ref{seccs} satisfy the requirements established in \cite{Zel17}. Considering also that the bi-orthogonality permits to operate the non-Hermitian oscillators as in the conventional Hermitian case, it may be shown that the corresponding $P$-representation is proportional to the delta distribution. Nevertheless, the latter means the existence of a classical analog only if the states of Section~\ref{seccs} are fully coherent.

Instead of investigating whether our coherent states lack second (or higher) order coherence, we shall use some of the techniques described in Appendix~\ref{ApB} to study their classicalness. Namely, we are going to study the corresponding variances, Mandel parameter, Wigner distribution and purity. 

\subsection{Nonclassical optimized binomial states}

Let us investigate the properties of the distorted quadratures (\ref{distquad}) in terms of the optimized binomial states (\ref{finite})-(\ref{finite2}). The variances $(\Delta X_w)^2$ and $(\Delta P_w)^2$ may be expressed in the form (\ref{c1}) of Appendix~\ref{ApB}:
\begin{equation}
(\Delta X_w)^2 = \tfrac14 \vert \langle I_w \rangle_b \vert +U_1, \quad (\Delta P_w)^2 = \tfrac14 \vert \langle I_w \rangle_b \vert - U_2,
\label{deltas}
\end{equation}
where we have used (\ref{distquad2}) and $\langle I_{w}\rangle_{b}\equiv \langle\overline{\phi}_{b}\vert I_{w}\vert\phi_{b}\rangle=1+(w-1)\left( \delta_{r,1} \, c_{0}^{2}+\delta_{r,0} \, c_{1}^{2} \right)-\delta_{r,0} \, c_{0}^{2}$. The straightforward calculation gives $U_{1}= \frac{1}{2}\left( M_{1}+M_{2}\right)-M_{3}^{2}$ and $U_{2}=\frac{1}{2}\left(M_{1}-M_{2}\right)$, with

\begin{eqnarray}
&& M_1=\sum_{k=0}^{N-2} \Omega_{k,r}^{(1)} c_{k+2} c_k \sqrt{(k+r+w-1) (k+r+w)},\\
&& M_2= \sum_{k=0}^N (k+r+w-2) \Omega_{k,r}^{(2)} c_k^2,\quad
M_3=\sum_{k=0}^{N-1} \Omega_{k,r}^{(1)} c_{k+1} c_k \sqrt{(k+r+w-1)}.
\end{eqnarray}
In the above expressions $\Omega_{k,r}^{(1)}=1-\delta_{k,r} \delta_{r,0}$, and $\Omega_{k,r}^{(2)}= \Omega_{k,r}^{(1)} - \delta_{k,1} \delta_{r,0}-\delta_{k,0} \delta_{r,1}$. Hereafter we make $\theta_k=0$ in $c_k = \vert c_k^b \vert e^{i\theta_k}$. 

\begin{figure}[htb]
\centering
\subfigure[$\omega =0$]{\includegraphics[width=0.25\textwidth]{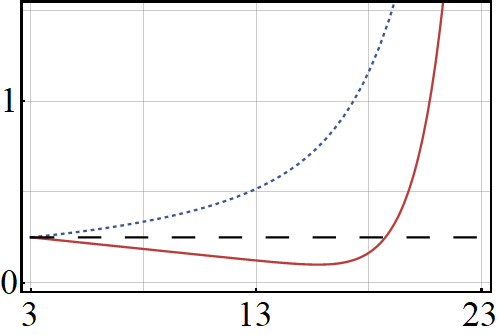}} 
\hskip4ex
\subfigure[$\omega =1$]{\includegraphics[width=0.25\textwidth]{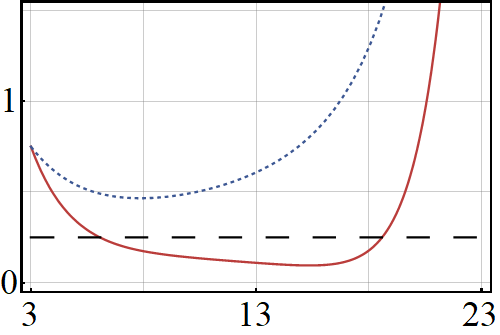}}

\caption{\footnotesize 
(Color online) The variances (\ref{deltas}) and expectation value $\tfrac14 \langle I_w \rangle_b$ of the optimized binomial states (\ref{finite}).  The functions $(\Delta X_w)^2$ and $(\Delta P_w)^2 $ are  depicted in solid-red and dotted-blue, respectively; the expectation value is in dashed-black. We have used $N=10$, $r=2$, and the indicated values of $\omega$. The horizontal axis refers to the allowed mean energy ${\cal E}_b$ introduced in Eqs.~(\ref{enbin})-(\ref{bound}).
}
\label{FigVB}
\end{figure}

It may be shown that there is always a subset of the mean energies (\ref{enbin})-(\ref{bound}) for which $(\Delta X_w)^2 <  \tfrac14 \langle I_w \rangle_b$. Thus, it is always possible to find a mean energy ${\cal E}_b$ producing the squeezing of the distorted quadrature $X_w$. The latter is illustrated in Fig.~\ref{FigVB} for $r=2$, $N=10$ and two different values of $w$. In turn, the squeezing of $P_w$ occurs for $r=0$ only, and it is not as strong as the one suffered by $X_w$ since $(\Delta P_w)^2$ is just slightly larger than $\tfrac14 \langle I_w \rangle_b$. In general, the optimized binomial states (\ref{finite}) are of minimal uncertainty (i.e., they are classical states) only at the lowest mean energy for either (i) $r=0$ and any value of $w$, or (ii) $r=2$ and $w=0$.

\begin{figure}[htb]
\centering
\subfigure[]{\includegraphics[width=0.3\textwidth]{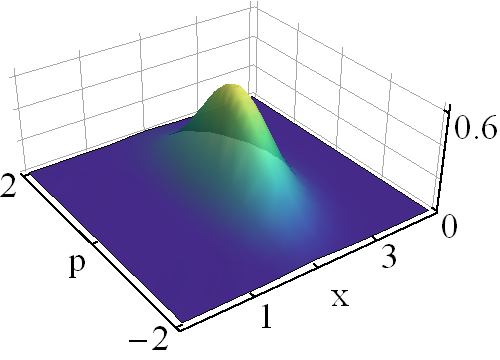}} 
\hspace{.3ex}
\subfigure[]{\includegraphics[width=0.3\textwidth]{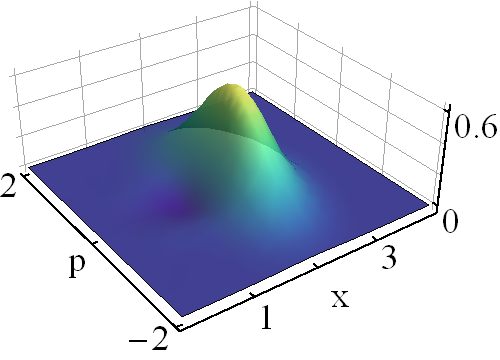}}
\hspace{.3ex}
\centering
\subfigure[]{\includegraphics[width=0.3\textwidth]{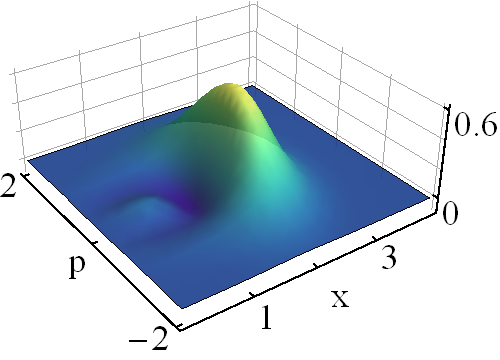}}

\caption{\footnotesize 
Wigner distribution of the optimized binomial sates (\ref{opclas}). The panel is constructed with (a) $p =0.5$, $r=0$, (b) $p =0.4$, $r=1$, and (c) $p =0.3$, $r=2$. In all cases $N=10$ and ${\cal E}_b= 9$.}
\label{FigWB}
\end{figure}

In the quantum oscillator limit (\ref{osc1}) one has
\begin{equation}
\phi_{b}^{osc} (x) = \left. \lim_{\vert \gamma \vert \rightarrow  \infty}  \phi_b (x) \right \vert_{\lambda=0}=
\sum_{k=0}^{N}
\vert c_k^b \vert e^{i\theta_k} \varphi_{k+r} (x).
\label{opclas}
\end{equation}
Clearly, we recover the conventional binomial state for $r=0$. This state is squeezed in the quadrature $\hat x$, for $N=10$ and $p =0.5$, as shown in the Wigner distribution of Fig.~\ref{FigWB}(a). If $r \geq 1$, besides the squeezing of $\hat x$, the Wigner distribution exhibits negative values in different zones of the phase-space, see Figs.~\ref{FigWB}(b) and  \ref{FigWB}(c). The latter results confirm that the optimized binomial states (\ref{opclas}) are stronger nonclassical than the conventional ones. 

\begin{figure}[htb]
\centering
\includegraphics[width=0.25\textwidth]{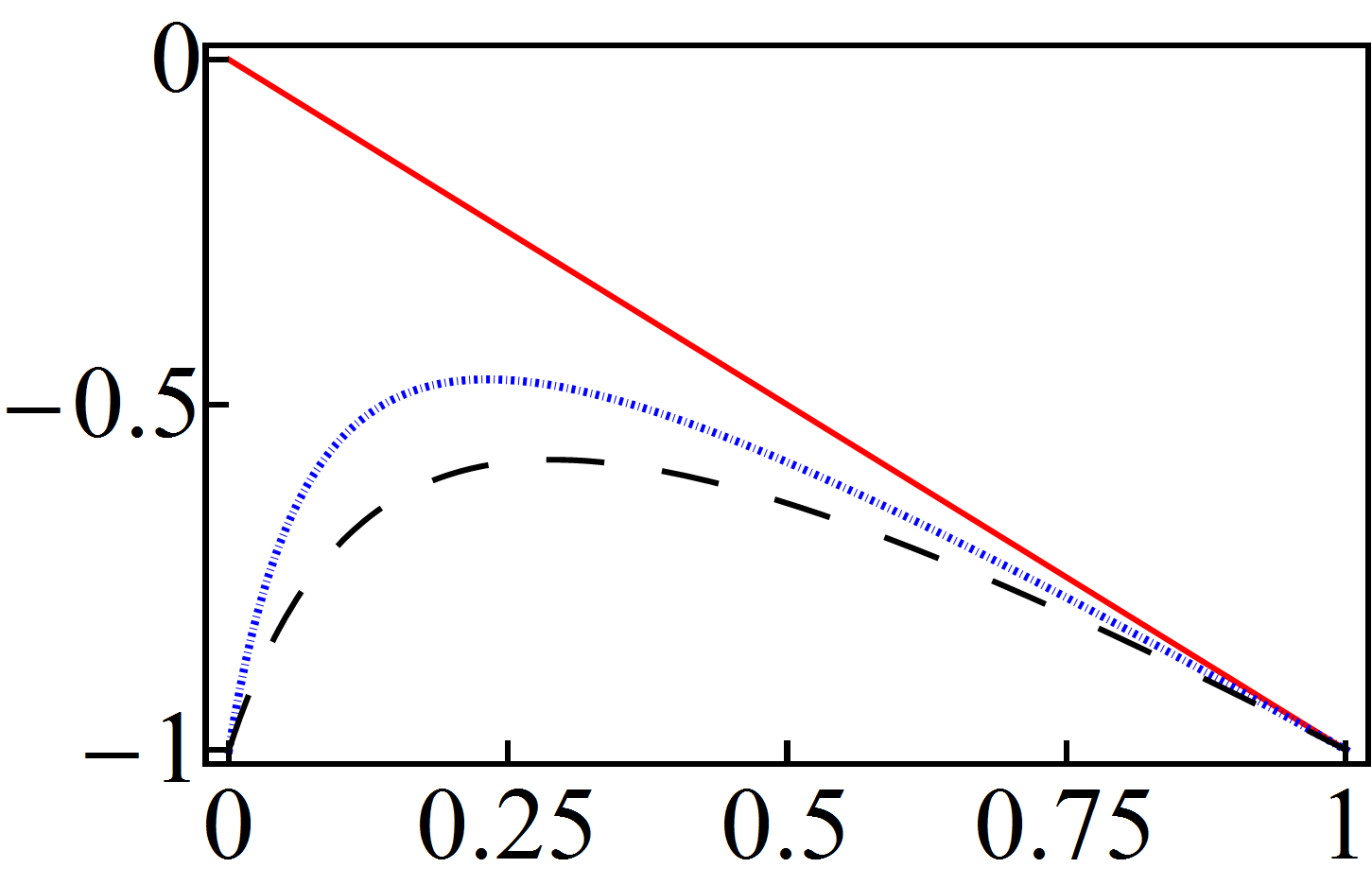}

\caption{\footnotesize 
(Color online) Mandel parameter $Q$ of the optimized binomial sates (\ref{opclas}) in terms of the probability $p$. The curves correspond to $N=10$ with $r=0$ (red-solid), $r=1$ (blue-dotted) and $r=2$ (black-dashed).}
\label{FigQB}
\end{figure}

To get more insights about the optimized binomial states (\ref{opclas}) we have depicted the related Mandel parameter $Q$, as a function of $p$, in Fig.~\ref{FigQB}. For $p =1$ the superposition (\ref{finite}) is reduced to the state $\varphi_{N+r}(x)$, so that all the curves shown in Fig.~\ref{FigQB} converge to $-1$ as $p \rightarrow 1$. In turn, for $r\neq 0$, all the curves take the value $-1$ at $p=0$ since only the Fock state $\varphi_{r\neq 0} (x)$ is included in the sum. Thus, for $r\neq 0$ and any $p$, the Mandel parameter $Q$ in this case is always negative. Then, the optimized binomial states (\ref{opclas}) are sub-Poissonian for $r\neq 0$, and thus do not admit any classical description. 

\begin{figure}[htb]
\centering
\includegraphics[width=0.25\textwidth]{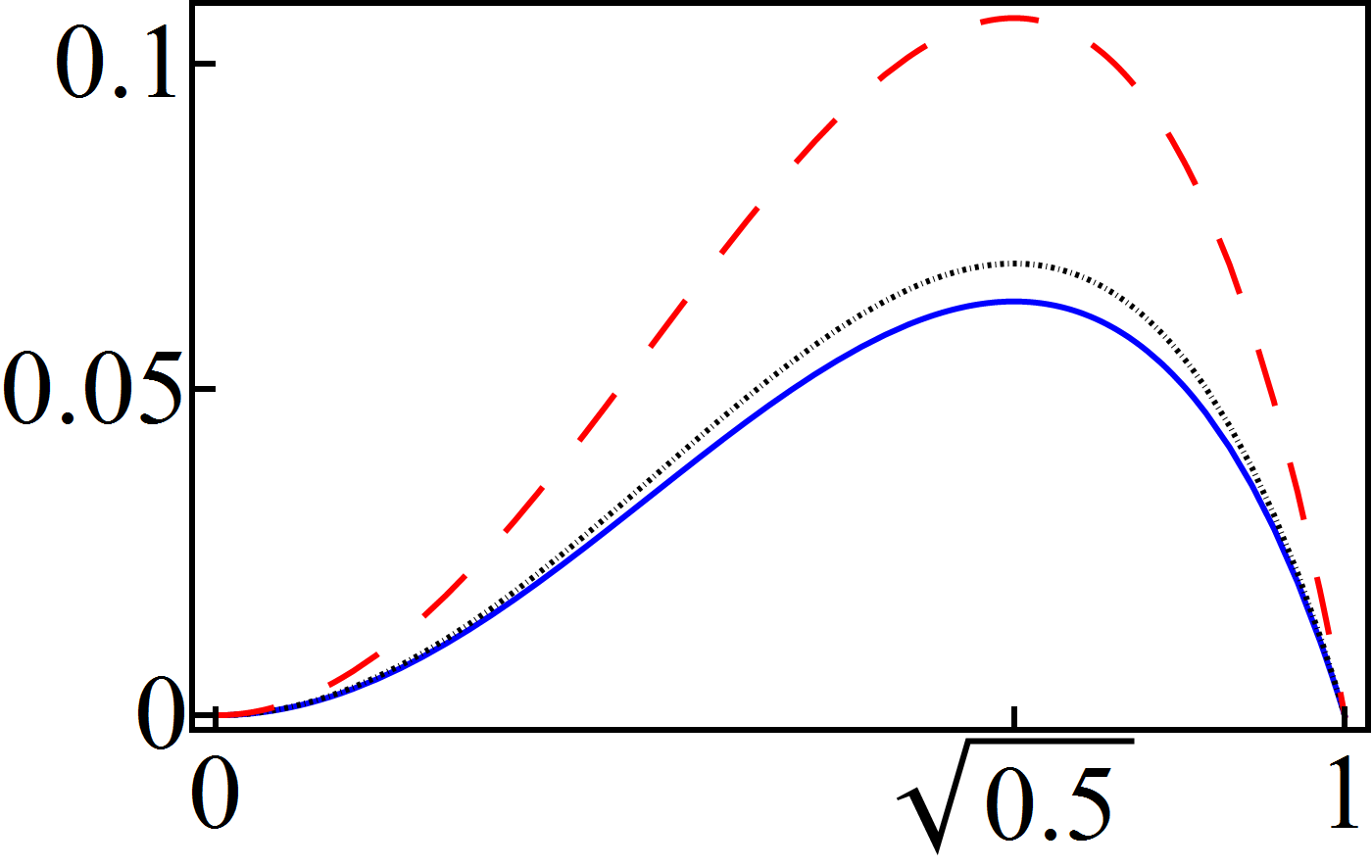}

\caption{\footnotesize 
(Color online) Purity $S_{L}$ of the optimized binomial states (\ref{opclas}) in terms of the transmission coefficient $T$ of a testing beam-splitter. In all cases $N=10$ and ${\cal E}_b=9$. The curves correspond to $p =0.5$, $r=0$ (solid-blue), $p =0.4$, $r=1$ (dotted-black), and $p =0.3$, $r=2$ (dashed-red).}
\label{FigPuB1}
\end{figure}

Let us apply also the beam-splitter technique (see Appendix~\ref{ApB}) to the states we are dealing with. We may study the purity (linear entropy) $S_L$ in terms of either the transmission coefficient $T$ of the beam-splitter or the probability $p$ that defines the superposition (\ref{opclas}). In the former case it is convenient to fix $N$ and ${\cal E}_b$ to parameterize the purity with $r$ and $p$. Then one finds that the purity is equal to zero for $T=0$ and $T=1$, see Fig.~\ref{FigPuB1}. However, it is important to remark that such results give no special information since they correspond to complete reflectance (i.e., the beam-splitter is indeed a perfect mirror) and complete transparency (no beam-splitter), so the output is respectively the vacuum and the state which was injected in input. In general, for other values of $T$, the purity reaches its maximum at $T=\sqrt{0.5}$, i.e., when the testing beam-splitter is $50/50$. 

\begin{figure}[htb]
\centering
\includegraphics[width=0.25\textwidth]{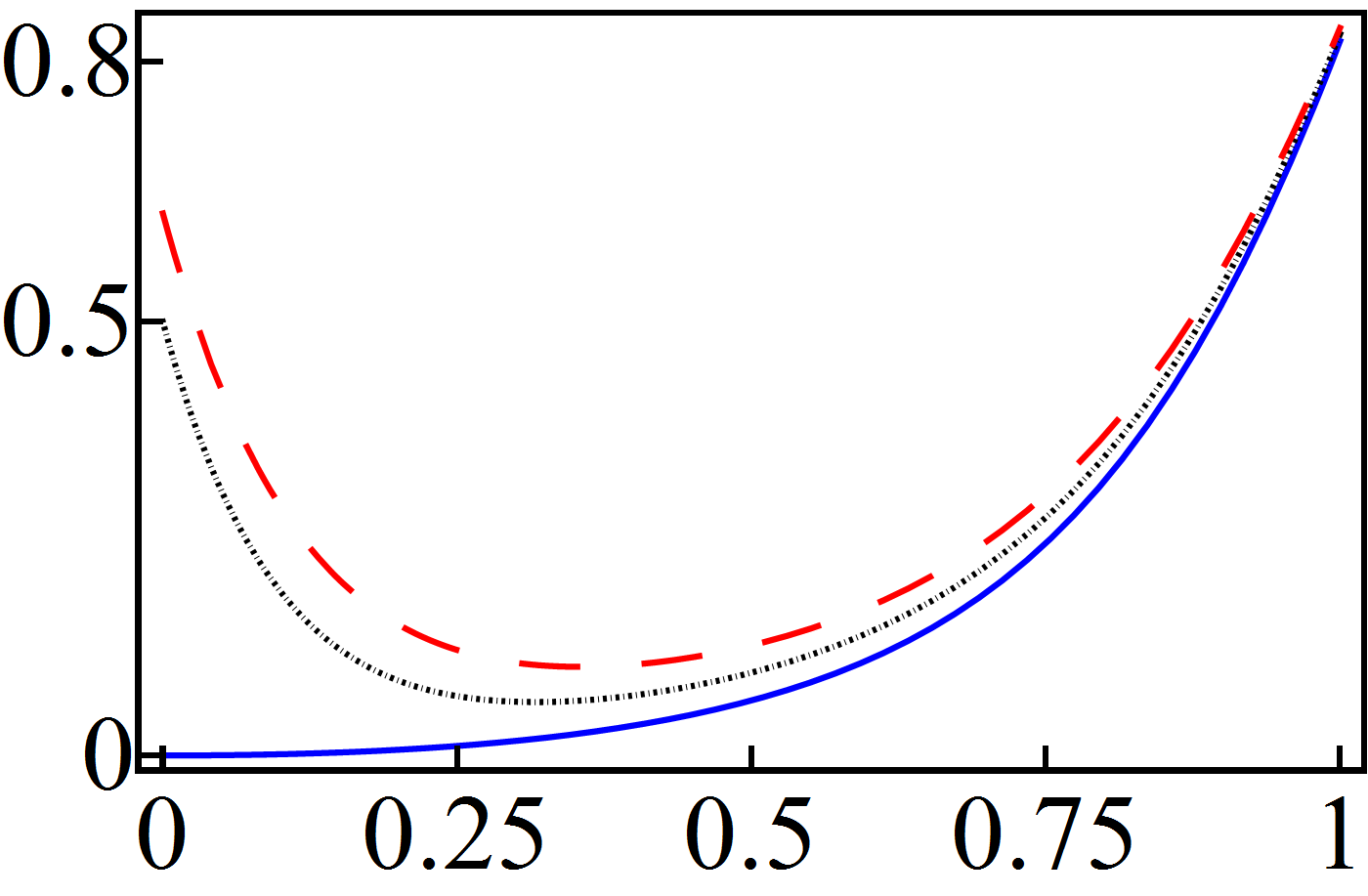}

\caption{\footnotesize 
(Color online) Purity $S_L$ of the optimized binomial states (\ref{opclas}) in terms of the probability $p$ with  $N=10$. We have used a $50/50$ beam-splitter ($T= \sqrt{0.5}$) for $r=0$ (solid-blue), $r=1$ (dotted-black) and $r=2$ (dashed-red). Compare with Fig.~\ref{FigPuB1}.
}
\label{FigPuB2}
\end{figure}

To analyze $S_L$ in terms of the probability $p$ we must fix $N$ and $T$. Then, with exception of the purity for $r=0$, we find $S_L >0$ for any value of $p$ (see Fig.~\ref{FigPuB2}). The latter confirms, once again, that the optimized binomial states $\phi_b (x)$ are nonclassical. Concerning the case $r=0$, the purity is equal to zero at $p=0$ only. Thus, the superposition (\ref{opclas}) only includes the vacuum state $\varphi_0(x)$, which is clearly classical.

\subsubsection{Nonclassical optimized Poisson states}

\begin{figure}[htb]

\centering
\subfigure[$\omega =0$, $r=1$]{\includegraphics[width=0.25\textwidth]{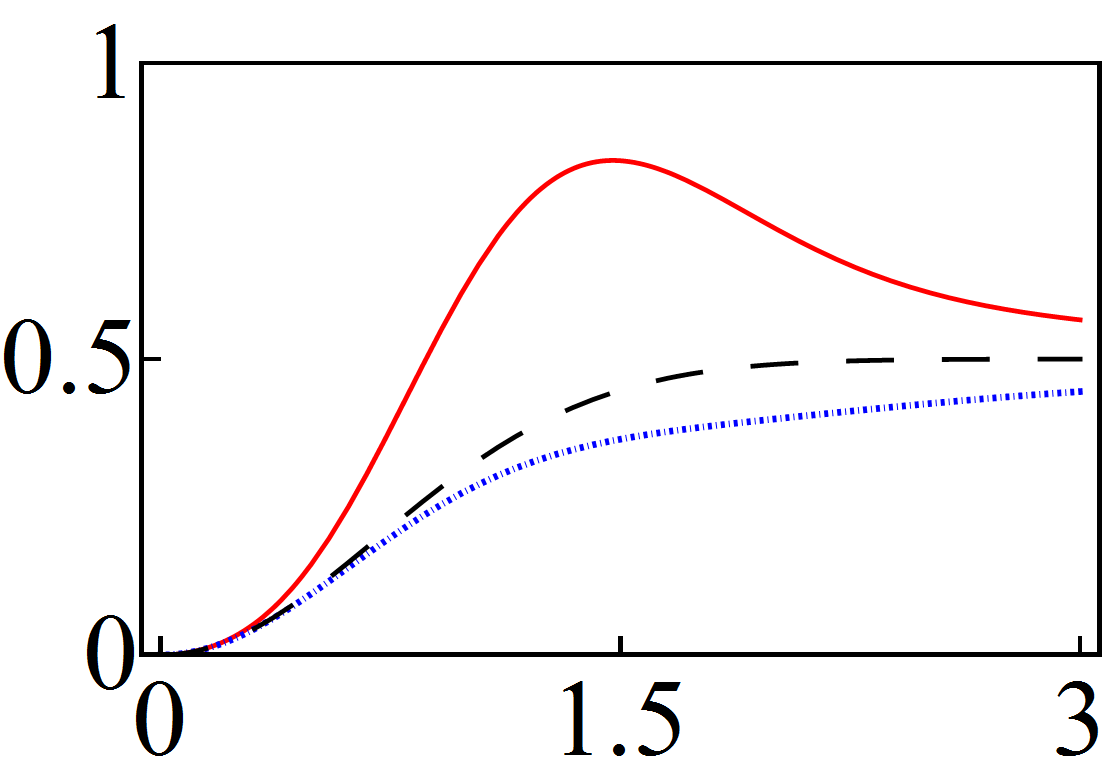}} 
\hskip4ex
\subfigure[$\omega =1$, $r=2$]{\includegraphics[width=0.25\textwidth]{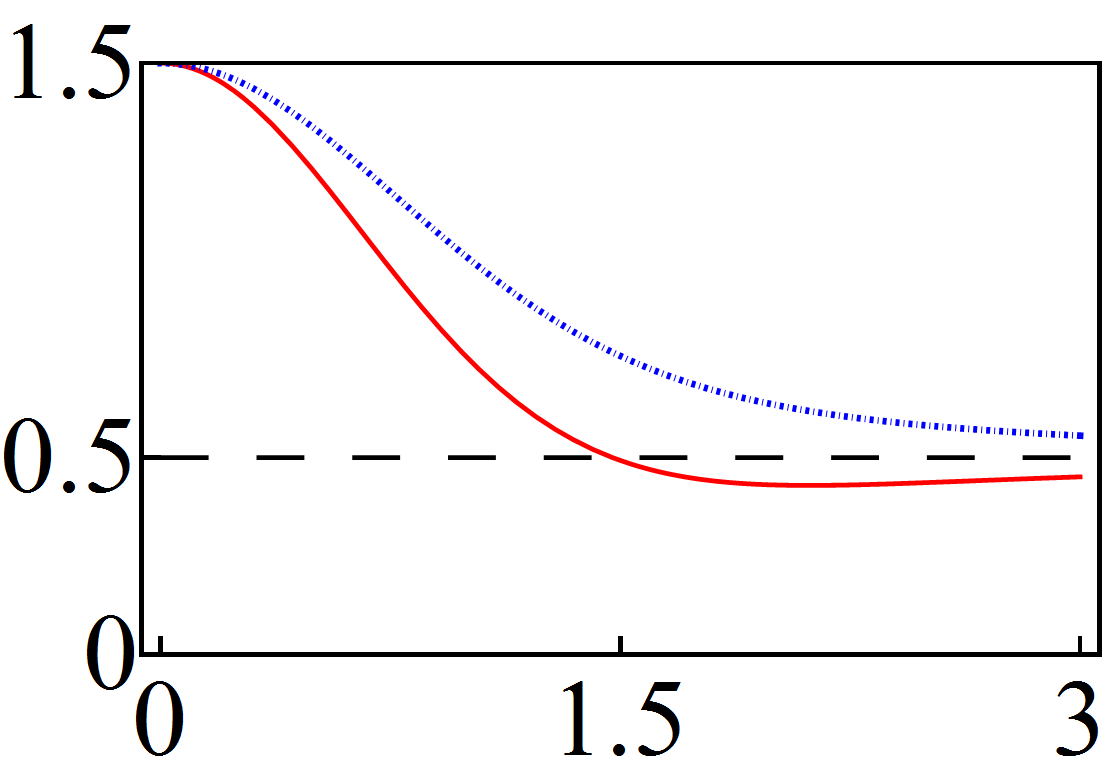}}

\caption{\footnotesize 
(Color online) The variances $(\Delta X_w)^2$ and $(\Delta P_w)^2$, in solid-red and dotted-blue, respectively, of the optimized Poisson states (\ref{poisson}) for the indicated values of $\omega$ and $r$. The minimum uncertainty (dashed-black) is included as a reference.}
\label{FigVP}
\end{figure}

It may be shown that the optimized Poisson states (\ref{poisson}) are minimum uncertainty states for concrete values of the parameters like either $r=1, \omega=1$, or $r=2,\omega=0$. In any case, they become states of minimum uncertainty as $\vert z \vert \rightarrow  \infty$. For other values of the parameters either $(\Delta X_w)^2$ or $(\Delta P_w)^2$ is squeezed, see Fig.~\ref{FigVP}.

\begin{figure}[htb]
\centering
\includegraphics[width=0.25\textwidth]{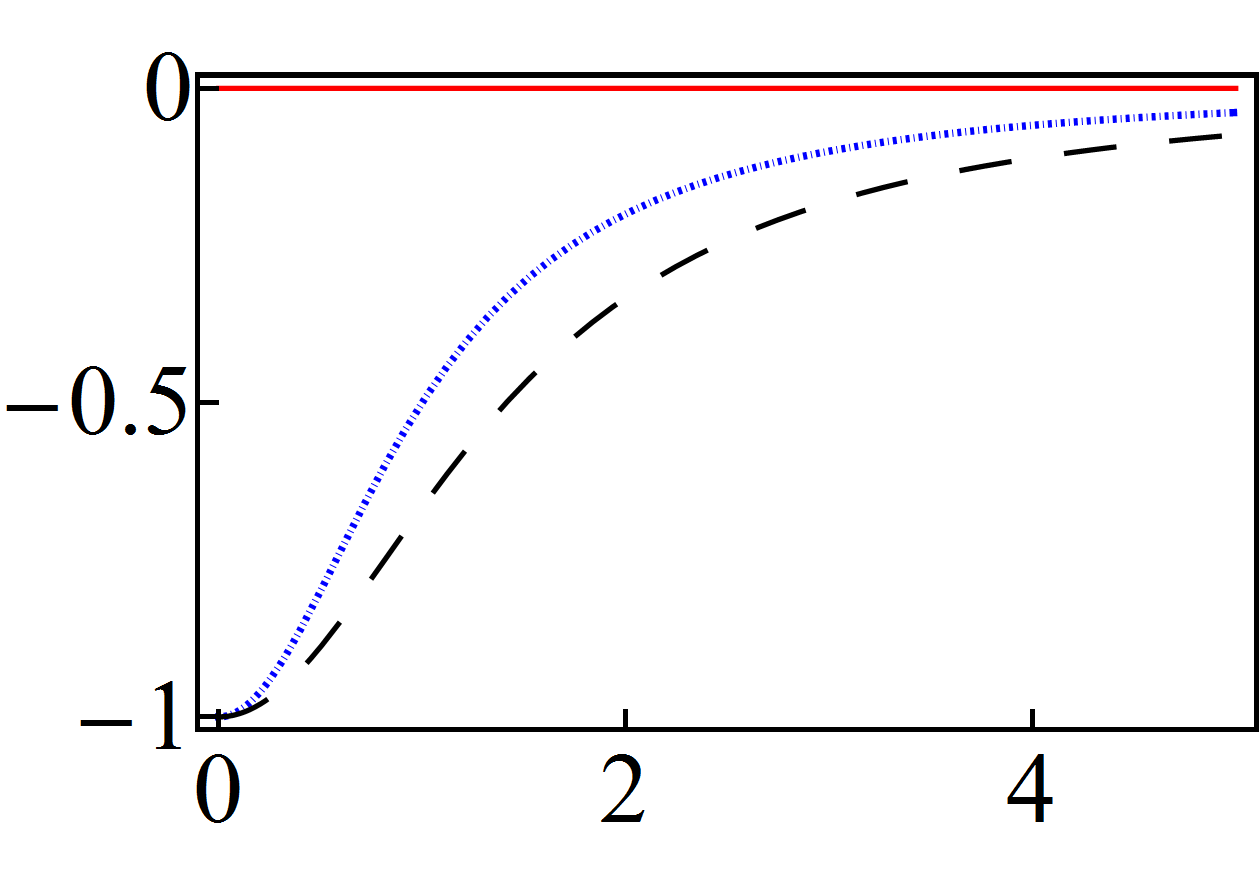}

\caption{\footnotesize 
(Color online) Mandel parameter $Q$ of the optimized Poisson states (\ref{opclas2}) as a function of $\vert z \vert$. The curves correspond to $r=0$ (solid-red), $r=1$ (dotted-blue) and $r=2$ (dashed-black).}
\label{FigQP}
\end{figure}

In the quantum oscillator limit (\ref{osc1}), the superposition (\ref{poisson}) acquires the form of Eq.~(\ref{opclas2}). As indicated above, for $r=0$ this state is reduced to the conventional coherent state, so that the Mandel parameter is equal to zero, as expected. For other values of $r$, the optimized Poisson states (\ref{opclas2}) are sub-Poissonian. In fact, in Fig.~\ref{FigQP} we see that the Mandel parameter becomes  zero as $\vert z \vert \rightarrow \infty$, and it is equal to $-1$ for $r\neq 0$ at $z=0$ (i.e. at the lowest mean energy ${\cal E}_b=2r-1$). In other words, for $r\neq 0$ and finite values of $z$, the optimized Poisson states $\phi_P^{osc} (x)$ are nonclassical. 

\begin{figure}[htb]
\centering
\subfigure[$r=0$]{\includegraphics[width=0.3\textwidth]{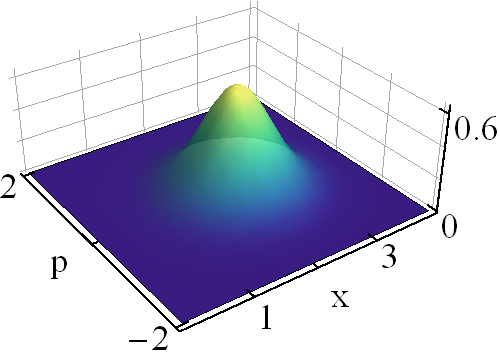}} 
\hspace{.4ex}
\subfigure[$r=1$]{\includegraphics[width=0.3\textwidth]{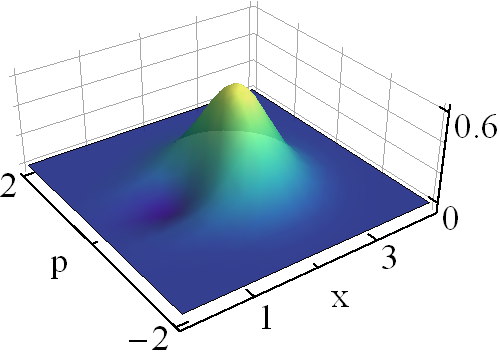}}
\hspace{.4ex}
\centering
\subfigure[$r=2$]{\includegraphics[width=0.3\textwidth]{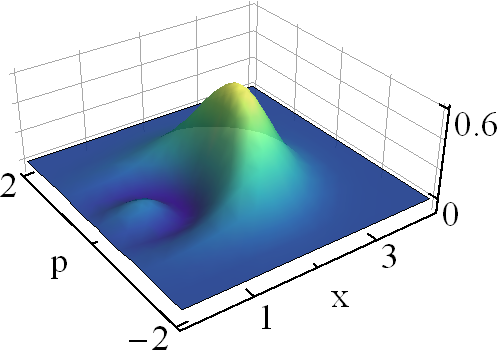}}
\hspace{.4ex}

\caption{\footnotesize 
Wigner distribution of the optimized Poisson states (\ref{opclas2}). The panel is constructed with $\vert z\vert =\sqrt{5}$ (a), $\vert z\vert =2$ (b), $\vert z\vert =\sqrt{3}$ (c), and the indicated values of $r$; in all cases ${\cal E}_b= 9$.}
\label{FigWP}
\end{figure}

For completeness, we show the Wigner function of $\phi_P^{osc} (x)$ in Fig.~\ref{FigWP}. Clearly, for $r=0$ one has a conventional (classical) coherent state. Other values of $r$ give rise to both, the squeezing of $\hat x$ and the negativity of the Wigner function in some regions of the phase-space.

$\bullet$ {\bf Photon added coherent states.} The expression (\ref{opclas2}) resembles the structure of the photon-added coherent states for the conventional harmonic oscillator \cite{Aga91}, written in Dirac notation as
\begin{equation}
\vert \alpha^{(pa)} \rangle = \sum_{k=0}^{+\infty} c_k^{(pa)} \vert \varphi_{k+r} \rangle,
\quad c_k^{(pa)}= \left[ \frac{(k+r)!}{ {}_1F_1(r+1,1,\vert \alpha \vert^2)\,r!} \right]^{1/2} \, \frac{\alpha^k}{k!}, \quad \alpha \in \mathbb C.
\label{padded}
\end{equation}
In both cases the superpositions are infinite and start with the eigenstate $\vert \varphi_r \rangle$. The main difference lies on the coefficients, for if we make $z=\alpha$, then
\begin{equation}
c_k^{(pa)}= \left[ \frac{(k+1)_{r}}{ {}_1F_1(r+1,1,\vert \alpha \vert^2)\,r!} \right]^{1/2}
e^{\frac{\vert \alpha \vert^2}{2}}  c_k^P,
\end{equation}
If $r=0$ we see that, up to the normalization constant $e^{-\vert \alpha \vert^2/2}$, the coefficients are the same. For other values of $r$ the superpositions (\ref{opclas2}) and (\ref{padded}) are different in general.

\begin{figure}[t]
\centering
\subfigure[$S_L(T)$, ${\cal E}_b=9$]{\includegraphics[width=0.25\textwidth]{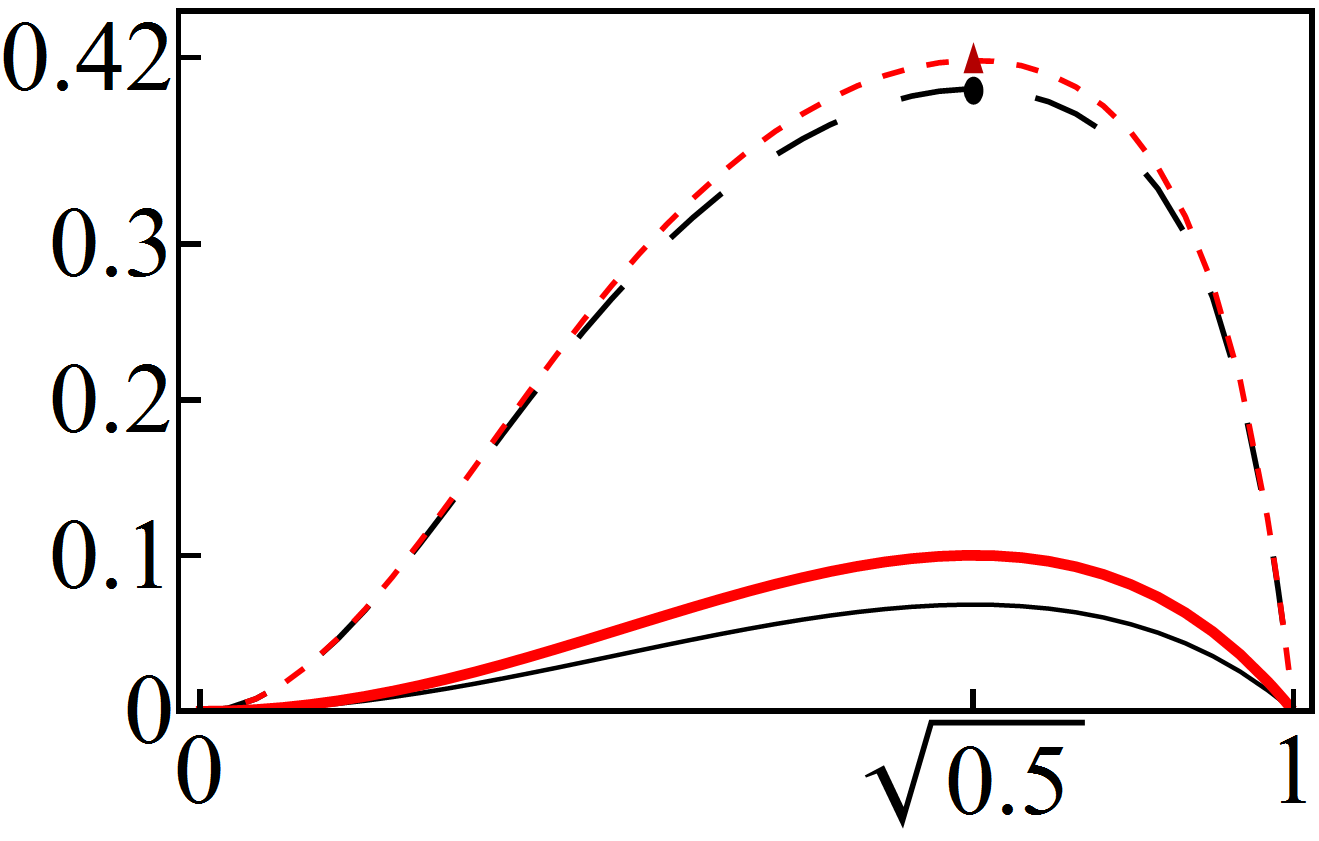}}
\hskip4ex
\subfigure[$S_L(\vert z \vert)$, $T=\sqrt{0.5}$]{\includegraphics[width=0.25\textwidth]{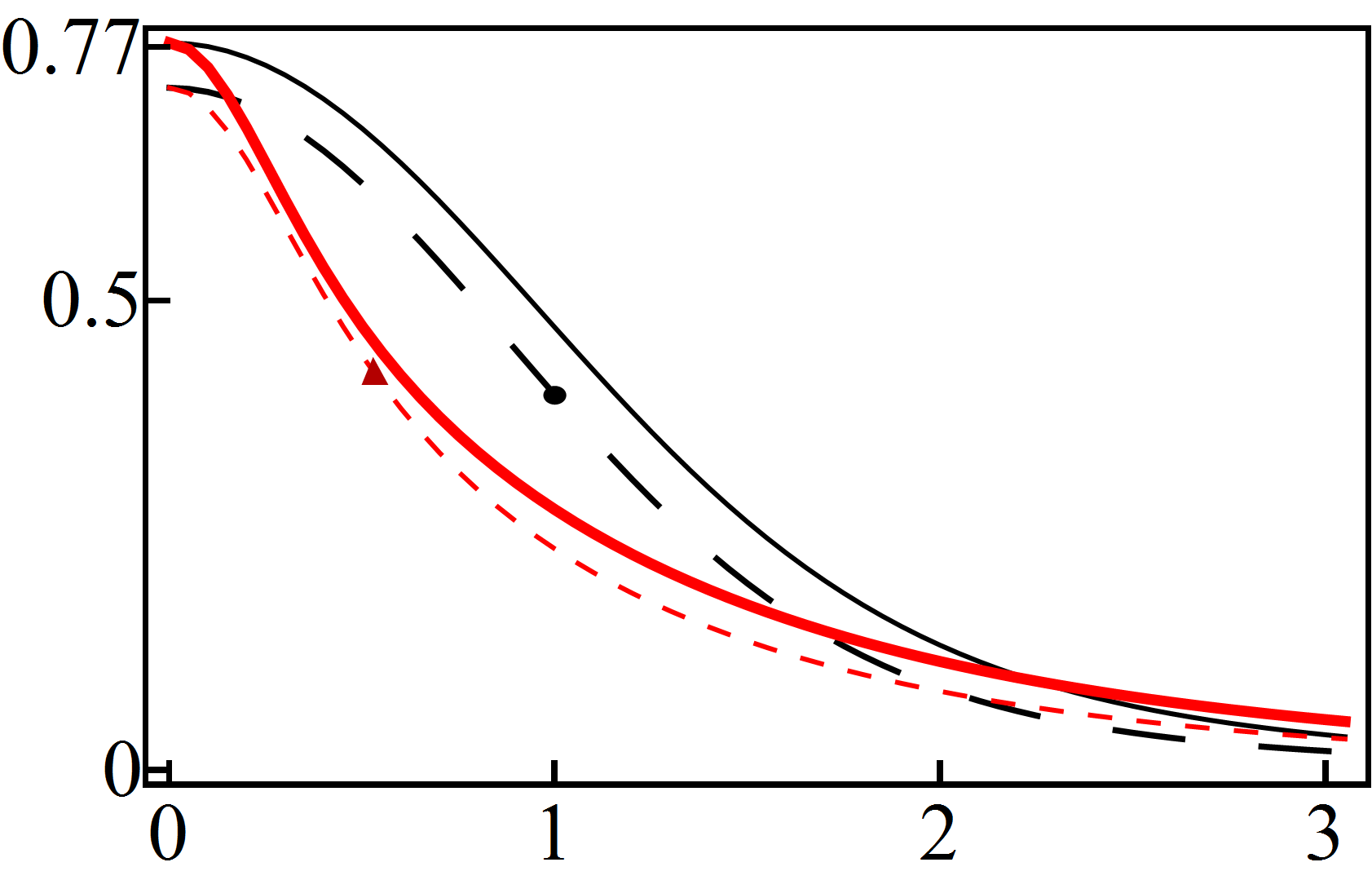}
}

\caption{\footnotesize 
(Color online) The purities of the photon-added coherent state (\ref{opclas2}) and the Poisson state (\ref{padded}) are respectively depicted in red and black. In (a) the mean energy is fixed,  ${\cal E}_b =9$, and the purities are plotted in terms of the transmission coefficient $T$ with  $r=2$ (thick-solid for (\ref{opclas2}) and solid for (\ref{padded})) and $r=4$ (shot-dashed for (\ref{opclas2}) and long-dashed for (\ref{padded})). In (b) the transmission coefficient is fixed, $T=\sqrt {0.5}$, and the purities are shown in terms of $z$ with $r=4$ (shot-dashed for (\ref{opclas2}) and long-dashed for (\ref{padded})) and $r=6$ (thick-solid for (\ref{opclas2}) and solid for (\ref{padded})). For comparison, in both figures the disk and triangle mark the purity evaluated at the same parameters.
}
\label{FigPuP}

\end{figure}

Let us compare $\vert \alpha^{(pa)} \rangle$ with $\vert \phi_P^{osc} \rangle$ in order to identify which one is stronger nonclassical. Using the beam-splitter technique, we find that the nonclassicality of these states depends on different parameters. For instance, in Fig.~\ref{FigPuP}(a) we have fixed the value of the mean energy as ${\cal E}_b=9$ and, after adjusting the parameters $z$, $\alpha$, and $r$, the purities $S_L(\phi_P^{osc})$ and  $S_L(\alpha^{(pa)})$ have been depicted in terms of the transmission coefficient $T$. As a global property, we see that the purity is maximum at $T= \sqrt{0.5}$ in all the cases. Besides, we can appreciate that $S_L(\phi_P^{osc}) < S_L(\alpha^{(pa)})$. Thus, for the parameters used in the figure the state $\vert \alpha^{(pa)} \rangle$ is stronger nonclassical than $\vert \phi_P^{osc} \rangle$. The situation changes if we fix $T$, and adjust the other parameters to plot the purity $S_L$ in terms of $\vert z \vert$. Then $S_L(\phi_P^{osc}) > S_L(\alpha^{(pa)})$ holds for small values of $\vert z\vert$ and $\vert \alpha\vert$, and $S_L(\phi_P^{osc}) < S_L(\alpha^{(pa)})$ is valid for large values of such parameters. The latter is illustrated in Fig.~\ref{FigPuP}(b) for $T=\sqrt{0.5}$ and $\vert z\vert =\vert \alpha\vert$.

\subsection{Nonclassical natural coherent states}

\begin{figure}[htb]

\centering
\subfigure[$S_L(\vert z \vert)$]{\includegraphics[width=0.25\textwidth]{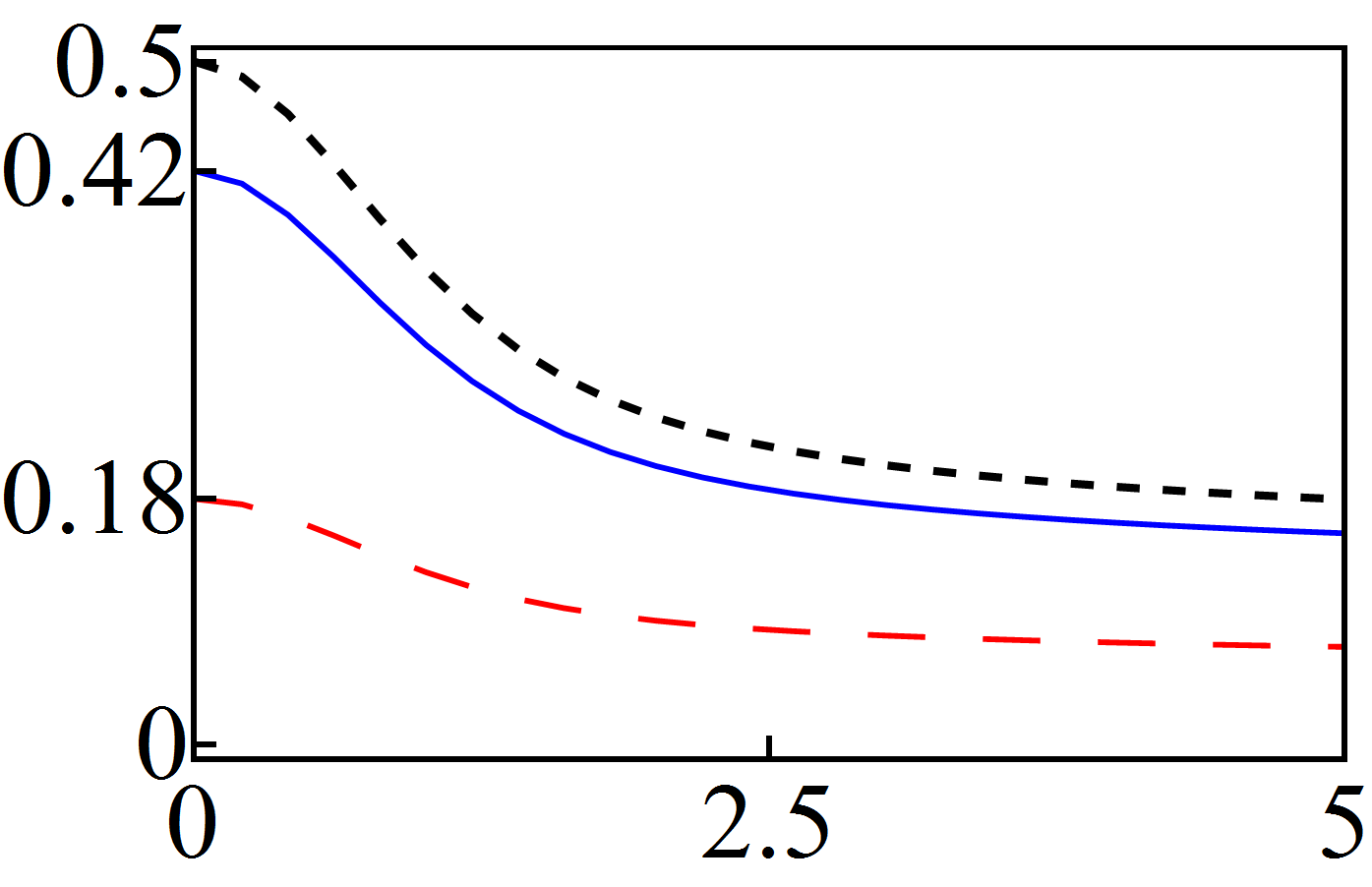}}
\hskip4ex
\subfigure[$Q(\vert z \vert)$]{\includegraphics[width=0.256\textwidth]{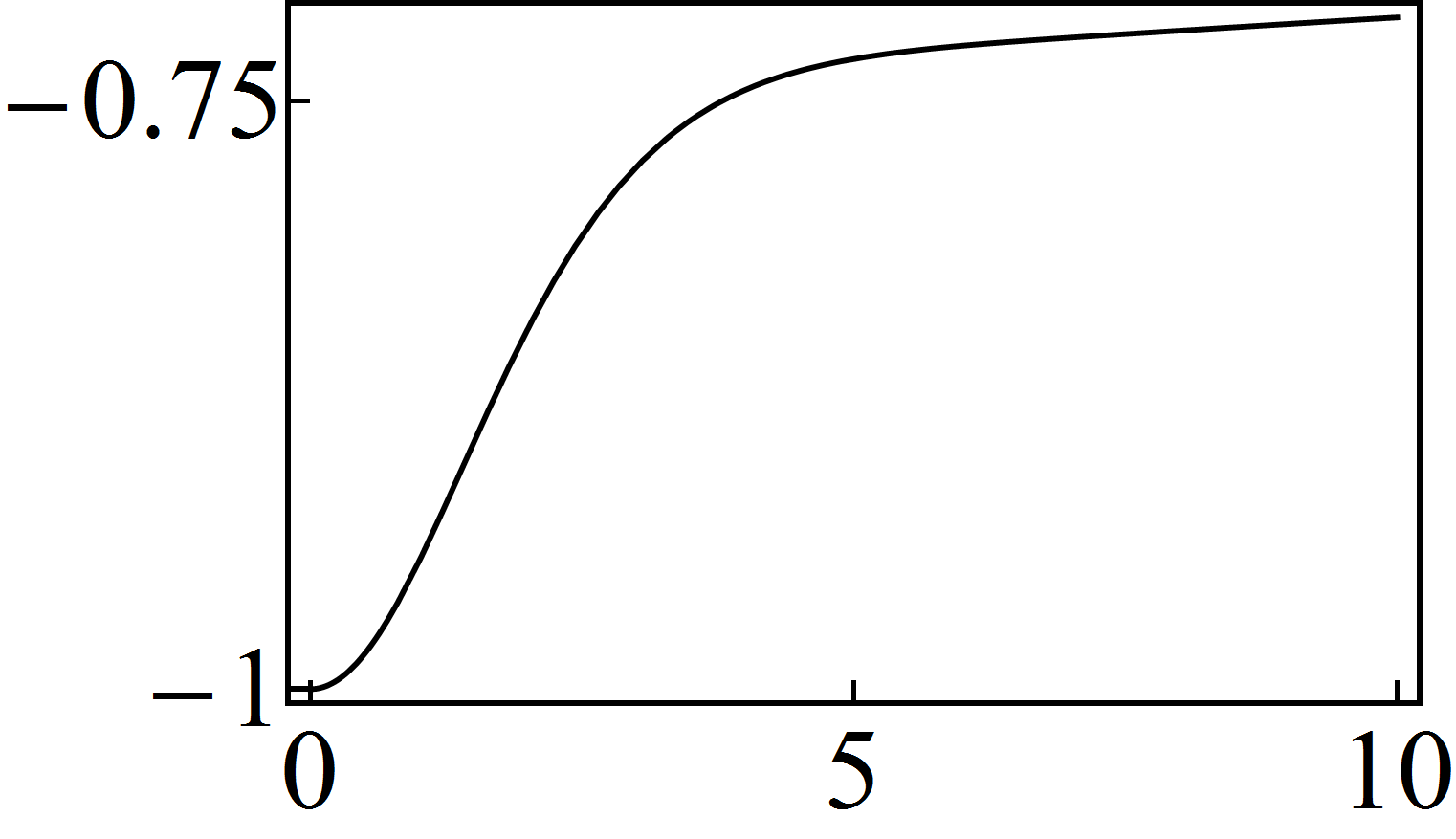}}

\caption{\footnotesize 
(Color online) (a) Purity of the natural coherent state $\vert z^{osc} \rangle$ in terms of $\vert z \vert$ for different values of the transmission coefficient: $T=\sqrt{0.1}$ (dashed-red), $T=\sqrt{0.3}$ (solid-blue) and $T=\sqrt{0.5}$ (dotted-black). (b) The corresponding Mandel parameter in terms of $z$.}
\label{FigLEN}
\end{figure}

By construction, the natural coherent states $\{ \vert \psi_0 \rangle, \vert z \rangle \}$ minimize the uncertainty relation  (\ref{quadnat2}) of Appendix~\ref{ApA} to investigate their properties in the oscillator limit (\ref{osc1}), we first calculate the purity $S_L$ using Eq.~(\ref{rho1}) of Appendix~\ref{ApB} with $r=1$, $c_k = c_k^{({\cal N})}$, and $N \rightarrow +\infty$ in the Fock basis $\vert  \varphi_k \rangle$. The result is depicted in Fig.~\ref{FigLEN}(a). As we can see,  this function is always different from zero and decreases as $\vert z \vert \rightarrow\infty$ for any value of the transmission coefficient $T$. That is, the states $\vert z^{osc} \rangle$ are nonclassical. The latter is confirmed by noticing that the Mandel parameter $Q$ is always negative, as it is  shown in Fig.~\ref{FigLEN}(b). 

\begin{figure}[htb]
\centering
\includegraphics[width=0.25\textwidth]{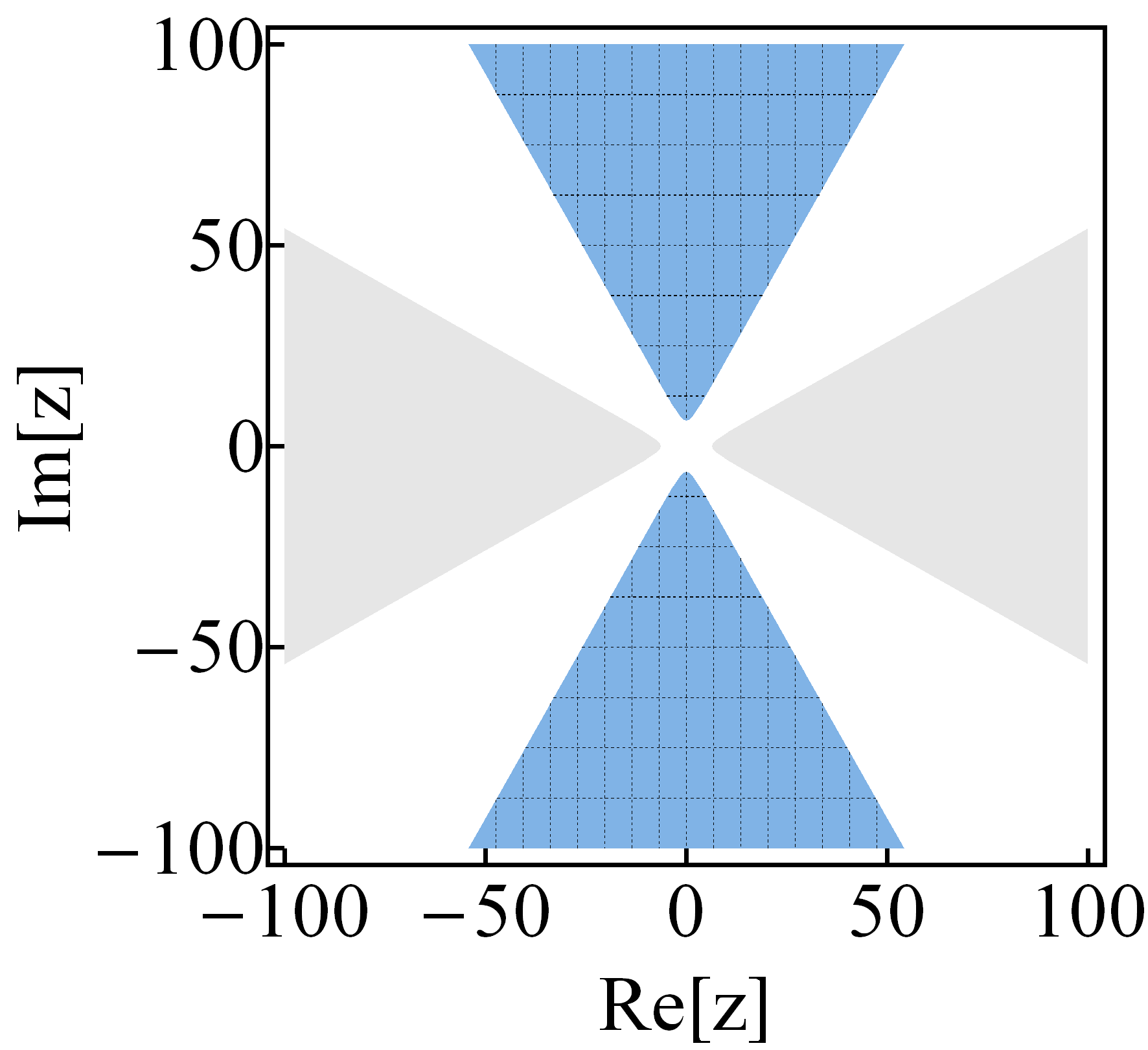}
\hskip4ex
\includegraphics[width=0.23\textwidth]{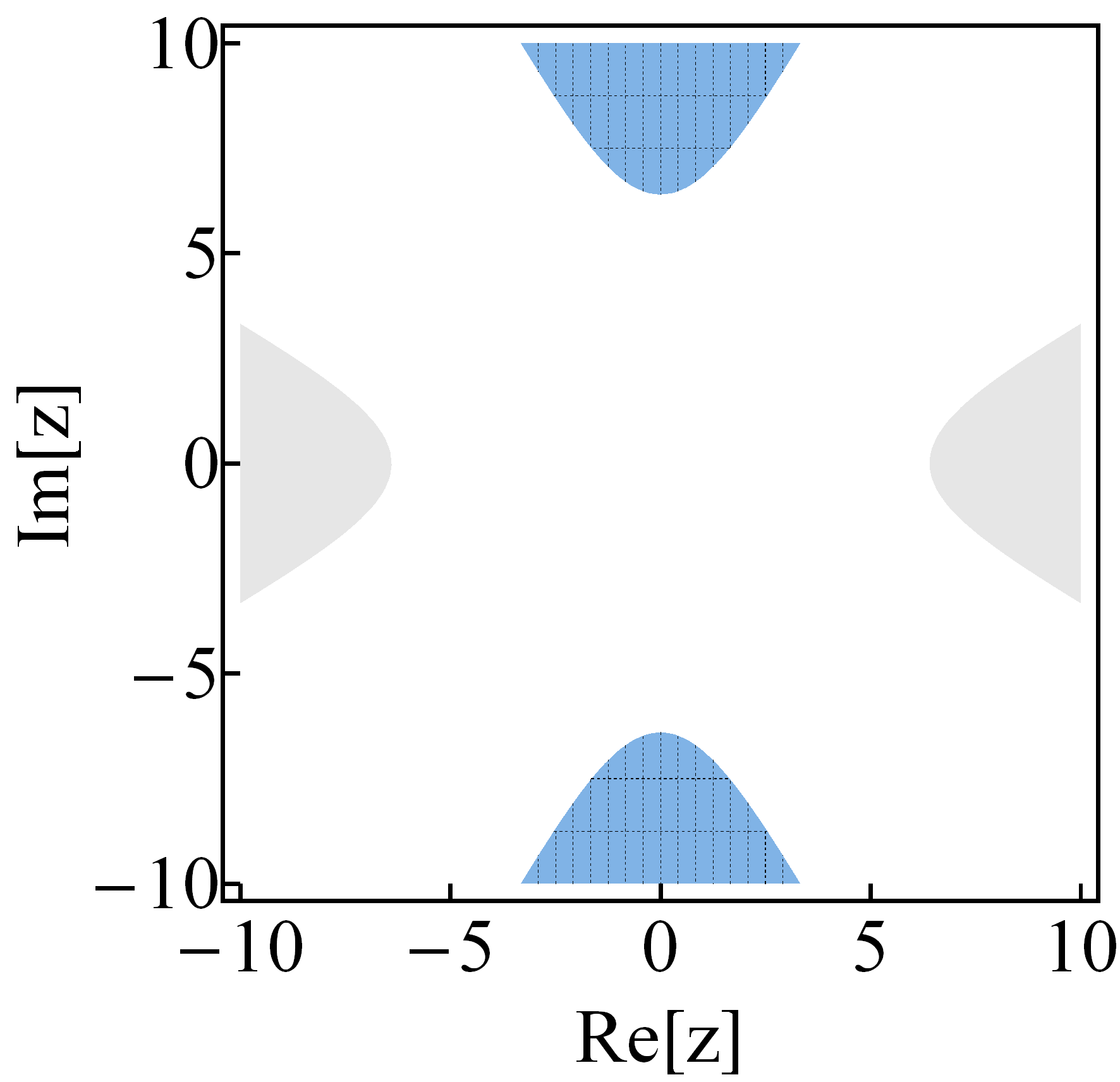}

\caption{\footnotesize 
(Color online) Regions of the complex plane where squeezing occurs. The grey (mesh, dashed blue) zones correspond to the values of $z\in \mathbb C$ for which $\hat x$ ($\hat p$) is squeezed. In turn, the grey and blue zones are centered at $\phi=0,\pi$ and $\phi=\tfrac{\pi}{2}, \tfrac{3\pi}{2}$, respectively.
}
\label{FigSqN}
\end{figure}

On the other hand, the variances of the physical quadratures $\hat x$ and $\hat p$ can be expressed in the form (\ref{c1}) of Appendix~\ref{ApB}, with $A=\hat x$, $B=\hat p$, and
\begin{equation}
\begin{array}{c}
U_1=1+ \left[ \mbox{Re}(z) \right]^2 f(z), \quad U_2=1+ \left[ \mbox{Im}(z) \right]^2 f(z),\\[2ex]
f(z)=\displaystyle\frac{{}_{0}F_{2}(2,3;\vert z \vert^2)}{{}_{0}F_{2}(1,2;\vert z \vert^2)} - 2 \left[ \frac{{}_{0}F_{2}(2,2;\vert z \vert^2)}{{}_{0}F_{2}(1,2;\vert z \vert^2)} \right]^{2}.
\end{array}
\end{equation}
In Fig.~\ref{FigSqN}, we show the regions of the complex $z$-plane where the squeezing of either $\hat x$ (grey zones) or $\hat p$ (mesh, dashed blue zones) occurs. Notice that the squeezing is maximum at the phase values $\phi=0,\pi$, and $\phi=\tfrac{\pi}2, \tfrac{3\pi}2$, respectively. The white zones are the regions of no-squeezing, they overrun four distinguishable areas defined along the phase values $\phi = \tfrac{k\pi}{4}$, $k=1,3,5,7$, and a circle of radius $\vert z \vert \approx 5$  that is centered at the origin. Comparing the Figs.~\ref{FigLEN} and \ref{FigSqN}, we see that the purity and the Mandel parameter give information that is complementary to that obtained from the variances of the quadratures. Namely, $S_L$ and $Q$ indicate strong nonclassicality for the states $\vert z^{osc} \rangle$ that satisfy the condition $\vert z \vert \lesssim 5$, where no squeezing is expected for neither $\hat x$ nor $\hat p$.

\begin{figure}[htb]
\centering
\subfigure[$z=0$]{\includegraphics[width=0.3\textwidth]{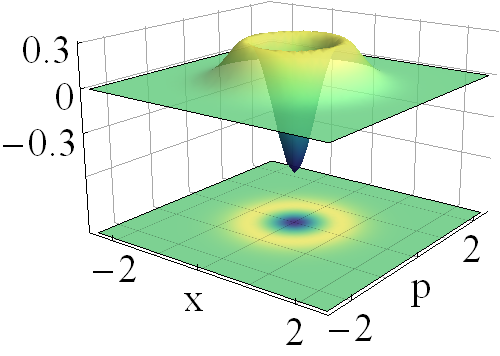}} 
\hskip2ex
\subfigure[$z=30$]{\includegraphics[width=0.3\textwidth]{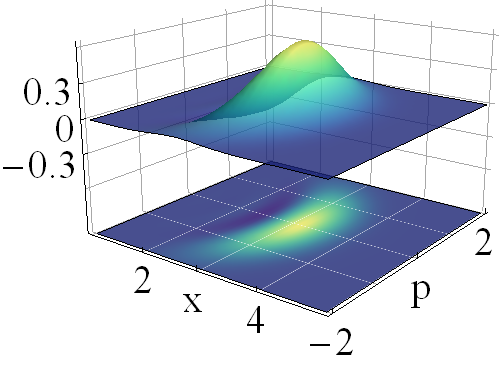}} 
\hskip2ex
\subfigure[$z=i30$]{\includegraphics[width=0.3\textwidth]{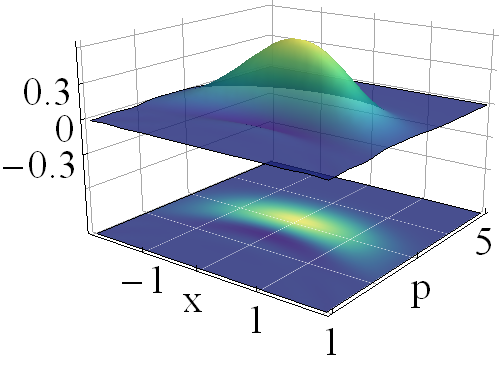}} 

\caption{\footnotesize 
Wigner distribution of the natural coherent sates $\vert z^{osc} \rangle$ for the indicated values of $z$.}
\label{FigWN}
\end{figure}

In Fig.~\ref{FigWN} we show the density plots of the Wigner function of $\vert z^{osc} \rangle$ for different values of $z$. In particular, for $z=0$ the natural coherent state $\vert z^{osc} \rangle$ is nonclassical because it coincides with the first excited state of the oscillator $\vert \varphi_1 \rangle$, as shown in Fig.~\ref{FigWN}(a). If $\vert z \vert >5$ and $\phi =0$ the quadrature $\hat x$ is squeezed; this is illustrated in Fig.~\ref{FigWN}(b) with $z=30$. In turn, for $z=i30$  one has the opposite result, $\hat p$ is squeezed; see Fig.~\ref{FigWN}(c).

\subsubsection{Even and odd natural coherent states}
\label{gcs11}

\begin{figure}[htb]
\centering{
\subfigure[$Q(\vert z \vert)$]{\includegraphics[width=0.3\textwidth]{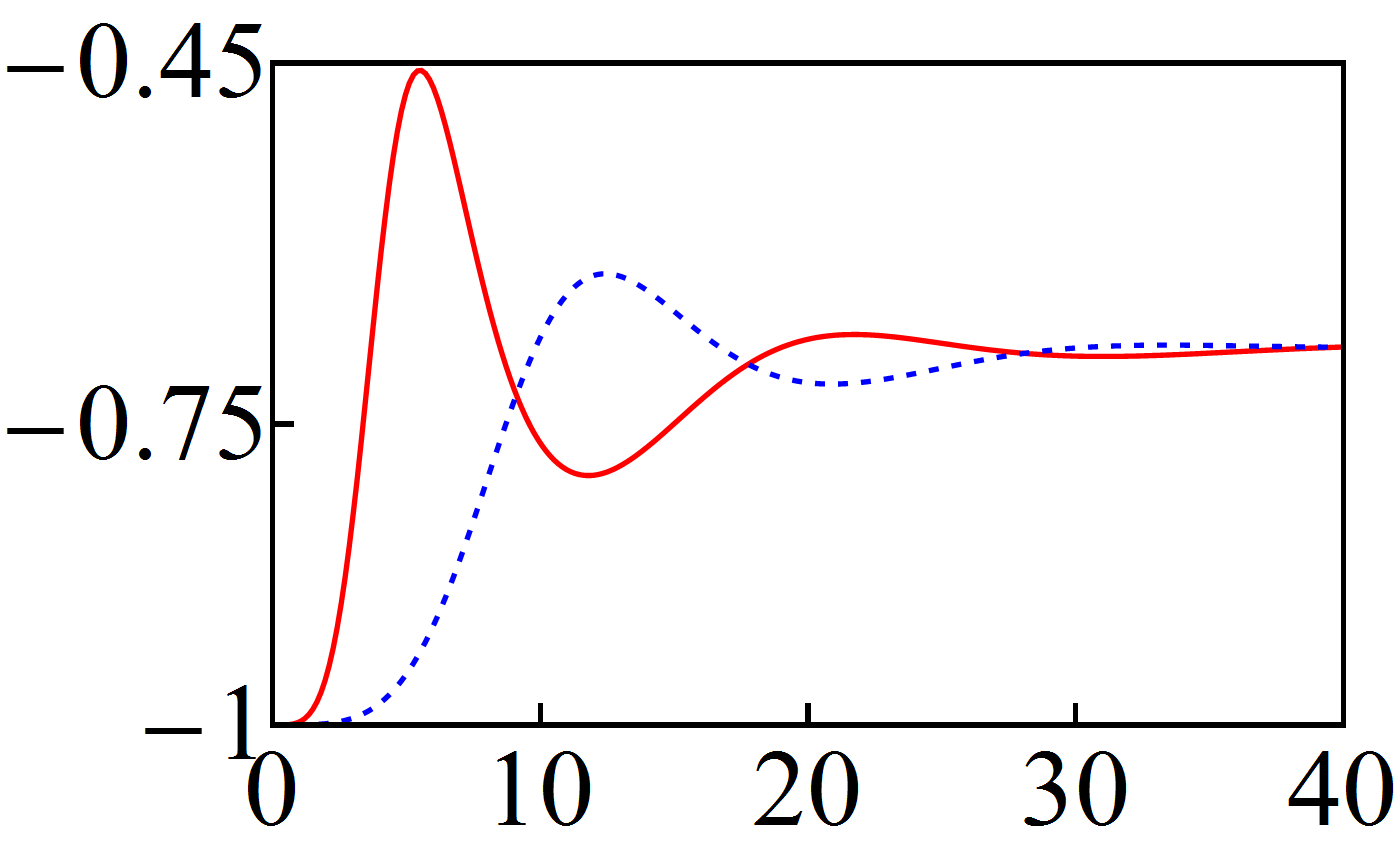}}
\hskip5ex
\subfigure[$z$-plane]{\includegraphics[width=0.25\textwidth]{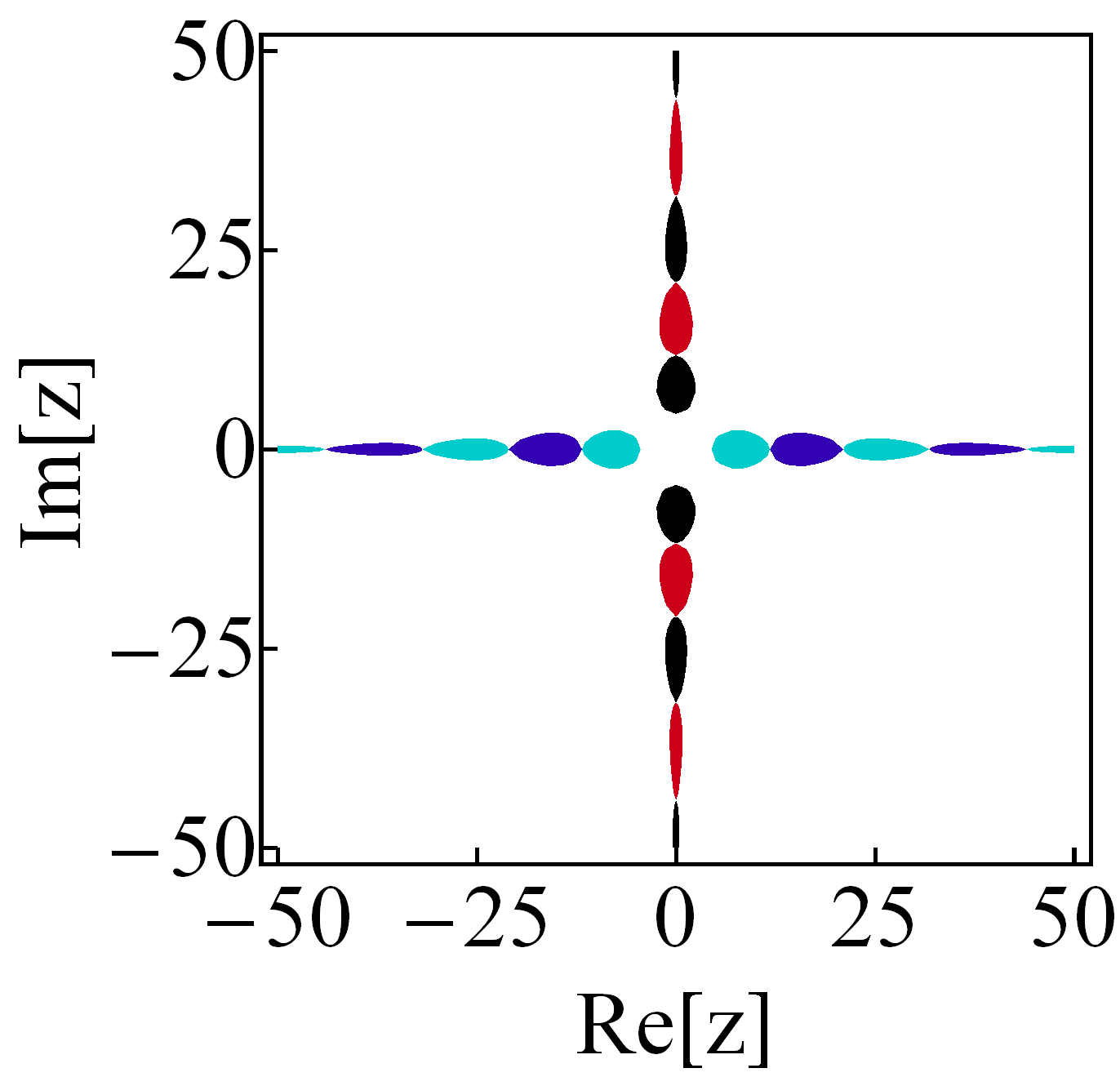}}
}
\caption{\footnotesize 
(Color online) (a) The Mandel parameter $Q$ of the even (solid-red) and odd (dahsed-blue) natural cat states (\ref{catn}) as a function of $\vert z \vert$. (b) Regions of the complex $z$-plane for which squeezing occurs. The red (even) and black (odd) regions correspond to the values of $z \in \mathbb C$ for which $P_{\cal N}$ is squeezed. Blue (even) and green (odd) regions correspond to the squeezing of $X_{\cal N}$.
}
\label{FigCatN1}
\end{figure}

We can use the natural coherent states (\ref{natural}) to construct additional nonclassical states. Following \cite{Dod74} we introduce the vectors
\begin{equation}
\vert z_{\pm} \rangle = \left[ 1 \pm \frac{ {}_0F_2 (1,2,-\vert z \vert^2)}{{}_0F_2 (1,2,\vert z \vert^2)}
\right]^{-1/2} \left( \frac{\vert z \rangle \pm \vert -z \rangle}{\sqrt 2} \right),
\label{catn}
\end{equation}
and call them {\em even} and {\em odd} natural cat states, respectively. The related Mandel parameter $Q$ is always negative, see Fig.~\ref{FigCatN1}(a), so that $\vert z_{\pm} \rangle$ are nonclassical. The variances of the natural quadratures (\ref{quadnat}) can be expressed in the form (\ref{c1}) of Appendix~\ref{ApB}, with $A= X_{\cal N}$, $B=P_{\cal N}$, and
\begin{equation}
\begin{array}{c}
U_1^{(\pm)} = 2 \left[ \displaystyle\frac{ \left[ \mbox{Re}(z) \right]^2 {}_0F_2(1,2; \vert z \vert^2) \mp  \left[ \mbox{Im}(z) \right]^2 {}_0F_2(1,2; -\vert z \vert^2)}{ {}_0F_2(1,2; \vert z \vert^2) \pm  {}_0F_2(1,2; -\vert z \vert^2) } \right],\\[3ex]
U_2^{(\pm)} = -2 \left[ \displaystyle\frac{ \left[ \mbox{Im}(z) \right]^2 {}_0F_2(1,2; \vert z \vert^2) \mp  \left[ \mbox{Re}(z) \right]^2 {}_0F_2(1,2; -\vert z \vert^2)}{ {}_0F_2(1,2; \vert z \vert^2) \pm  {}_0F_2(1,2; -\vert z \vert^2) } \right].
\end{array}
\end{equation}
In this case the quadratures are squeezed in very localized regions that are distributed along the real and imaginary axes of the complex $z$-plane. This is illustrated in Fig.~\ref{FigCatN1}(b), where we can appreciate a `discretization' of the eigenvalue $z$ as follows. The quadrature $X_{\cal N}$ is maximally squeezed along the real axis, in intervals $\vert z_k^{(j)} \vert \pm \ell_k^{(j)}$ that are defined by the points $\vert z_k^{(j)} \vert$, $k=1,2,\ldots$, with $\ell_k^{(j)} >0$, and $j$ denoting either even ($e$) or odd ($o$). For odd natural cats we make $j=o$, with $\vert z_1^{(o)} \vert$ defining the green zones that are closest to the origin in Fig.~\ref{FigCatN1}(b). For the even natural cats we write $j=e$, with $\vert z_1^{(e)} \vert$ defining the blue zones that are closest to the origin in the same figure. The intersection of the above intervals is empty, and they are such that the points $\vert z_k^{(j)} \vert$ interlace. That is $\vert z_k^{(o)} \vert < \vert z_k^{(e)} \vert < \vert z_{k+1}^{(o)} \vert < \vert z_{k+1}^{(e)} \vert< \cdots$. The phases constraining the green and blue zones are defined in intervals $\phi_k^{(j)} \pm d^{(j)}_k$, with $d^{(j)}_k >0$, that are shorter as $k$ increases. A similar description holds for the squeezing of $P_{\cal N}$ along the imaginary axis.

\begin{figure}[htb]
\centering
\subfigure[$\vert z \vert =0$]{\includegraphics[width=0.3\textwidth]{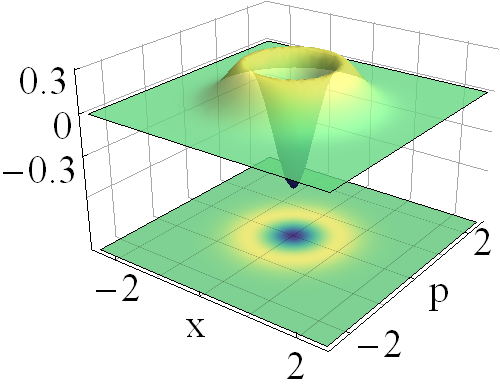}} 
\hskip1ex
\subfigure[$\vert z \vert =10$]{\includegraphics[width=0.3\textwidth]{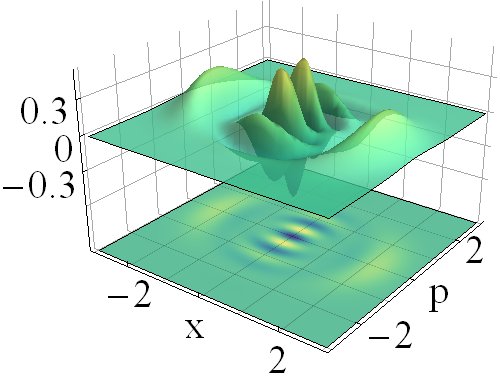}}
\hskip1ex
\centering
\subfigure[$\vert z \vert =30$]{\includegraphics[width=0.3\textwidth]{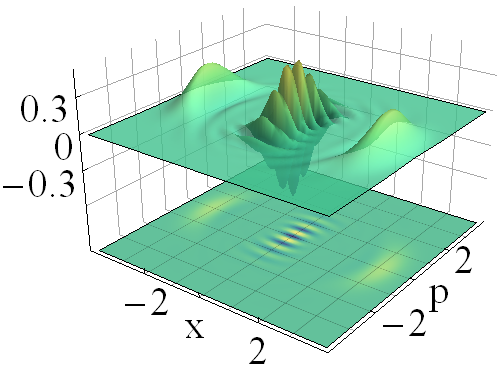}}
\hskip1ex

\centering
\subfigure[$\vert z \vert =0$]{\includegraphics[width=0.3\textwidth]{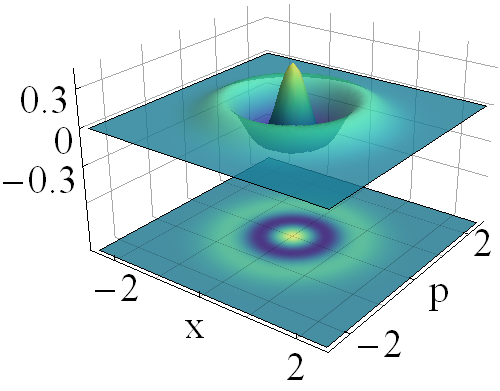}} 
\hskip1ex
\subfigure[$\vert z \vert =10$]{\includegraphics[width=0.3\textwidth]{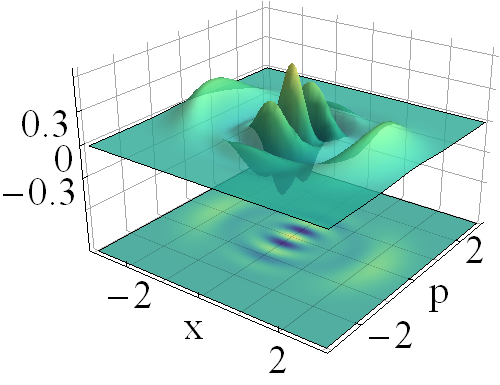}}
\hskip1ex
\centering
\subfigure[$\vert z \vert =30$]{\includegraphics[width=0.3\textwidth]{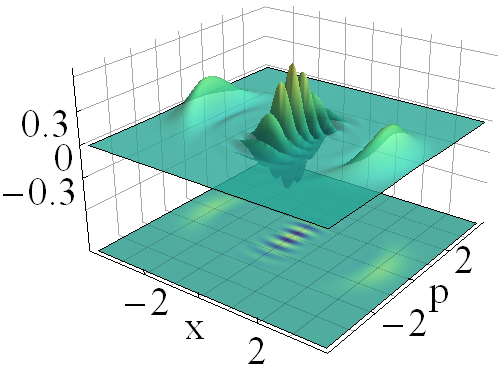}}
\hskip1ex

\caption{\footnotesize 
Wigner distributions of the even (upper row) and odd (lower row) natural cat states $\vert z_{\pm}^{osc} \rangle$ for $z= \vert z \vert e^{i\theta}$, with $\theta=0$ and the indicated values of $\vert z\vert$.
}
\label{FigCatN2}
\end{figure}

In the oscillator limit (\ref{osc1}), the natural cat states $\vert z_{\pm}^{osc} \rangle$ preserve their nonclassicality. In Fig.~\ref{FigCatN2} the Wigner function becomes negative in diverse zones of the complex $z$-plane. Such regions are less evident for $\vert z \vert \rightarrow \infty$.

\subsection{Nonclassical distorted coherent states}
\label{gcs2}

\begin{figure}[htb]
\centering
\subfigure[$Q(\vert z \vert)$]{\includegraphics[width=0.25\textwidth]{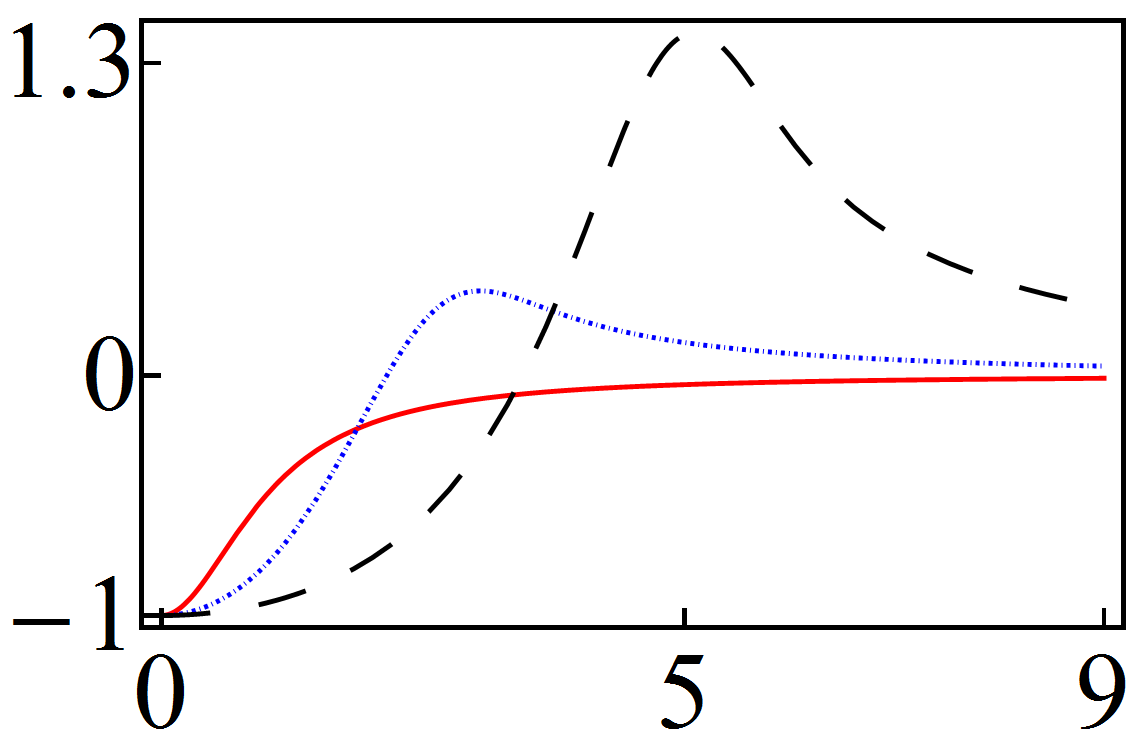}}
\hskip5ex
\subfigure[$z$-plane]{\includegraphics[width=0.2\textwidth]{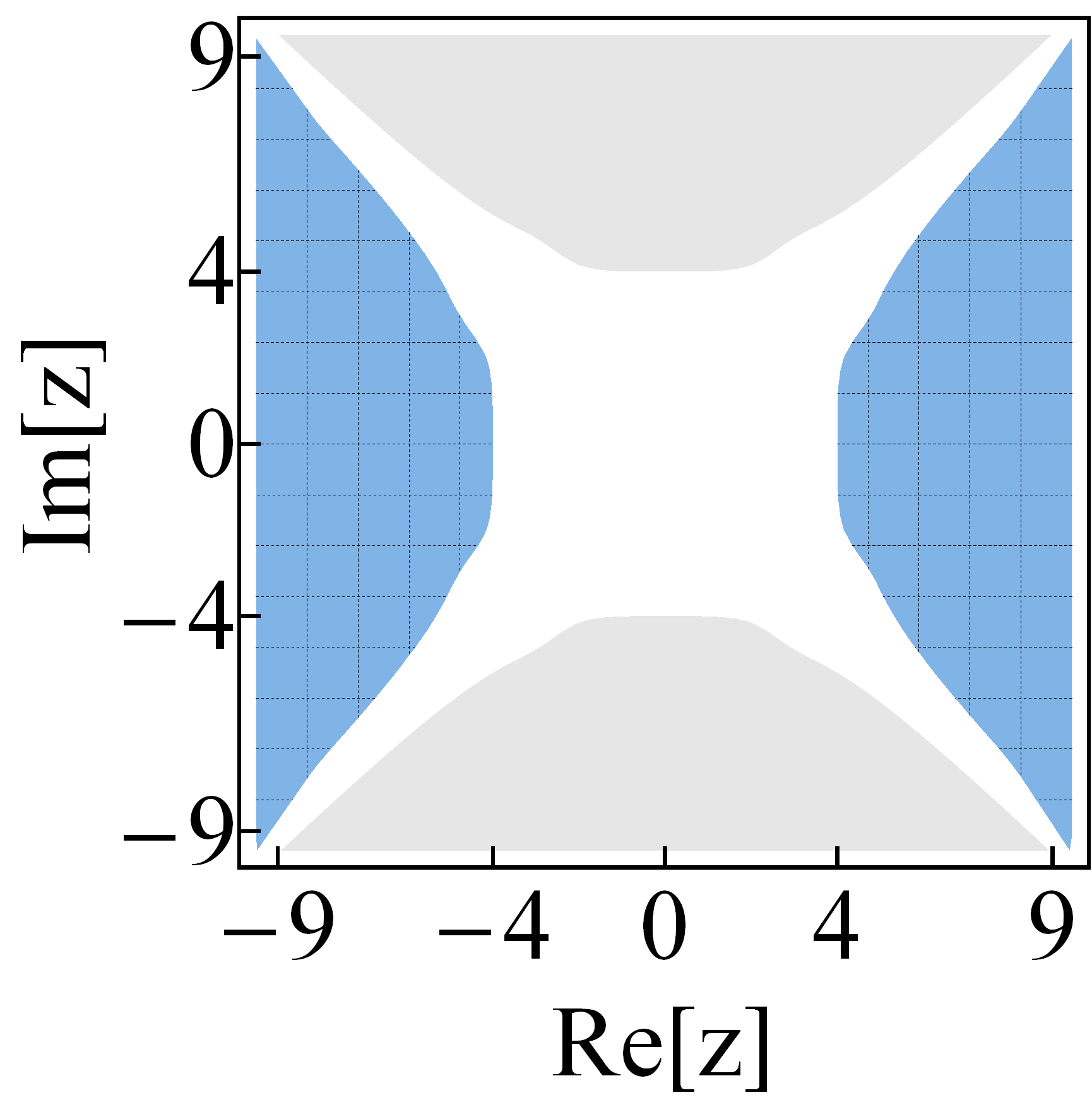}}

\caption{\footnotesize 
(Color online) The Mandel parameter $Q$ of the distorted coherent state $\vert z^{osc}, w\rangle$ is shown in (a) as a function of $\vert z\vert$ for $w=1$ (solid-red), $w=5$ (dotted-blue), and $w=20$ (dashed-black). In (b) we show the regions of squeezing in the complex $z$-plane for $w=20$. The code of colors is the same as in Fig.~\ref{FigSqN}(b).
}
\label{FigSqw}
\end{figure}

By construction, the distorted coherent states $\{ \vert \psi_0 \rangle, \vert z,w \rangle \}$ are minimal uncertainty states with respect to the uncertainty relation (\ref{distquad2}). In the oscillator limit, the Mandel parameter $Q$ depends on the parameter of distortion. In Fig.~\ref{FigSqw}(a), we see that the states $\vert z^{osc},w\rangle$ are sub-Poissonian for any value of $\vert z \vert$ whenever $w=1$. For arbitrary values of $w>1$, they are nonclassical only in the interval $(0, \vert z_w \vert)$. Here the value of $\vert z_w \vert$ is defined by $w$: it is larger if the value of $w$ is increased.

The variances of the physical quadratures $\hat x$ and $\hat p$ can be expressed in the form given by Eq.~(\ref{c1}) of Appendix~\ref{ApB}, with $A=\hat x$, $B=\hat p$, and
\begin{equation}
\begin{aligned}
U_1= \frac{\textnormal{Re}(z^2) f_2 ( \vert z \vert^2, \omega) + {}_1 F_1\left( 2, \omega , \vert z \vert^{2} \right) }{{}_1F_1(1,\omega; \vert z \vert^2)} - 2 \left[  \frac{\textnormal{Re}(z) f_1(z, \omega)}{{}_1F_1(1,\omega;\vert z \vert^2) } \right]^2, \\[2ex]
U_2= \frac{\textnormal{Re}(z^2) f_2(\vert z \vert^2, \omega) - {}_1 F_1\left( 2, \omega , \vert z \vert^{2} \right) }{{}_1F_1(1,\omega;\vert z \vert^2)}  + 2 \left[ \frac{\textnormal{Im}(z) f_1 (z, \omega)}{{}_1F_1(1,\omega;\vert z \vert^2)} \right]^2,
\end{aligned}
\end{equation} 
where
\begin{equation}
\begin{aligned}
f_1 (z, \omega)=\sum_{n=0}^{+\infty}\vert z \vert^{2n}\sqrt{\frac{n+2}{(\omega)_{n}(\omega)_{n+1}}}, \quad f_2 (\vert z \vert^2, \omega)=\sum_{n=0}^{+\infty}\vert z \vert^{2n}\sqrt{\frac{(n+2)(n+3)}{(\omega)_{n}(\omega)_{n+2}}}.
\end{aligned}
\label{efes}
\end{equation}
For $w \approx 1$, the quadratures $\hat x$ and $\hat p$ are squeezed along the real and imaginary axes, respectively. However, for larger values of $w$, they are squeezed along the imaginary and real axes. Thus, for $w>>1$ the squeezing is rotated by $\tfrac{\pi}2$, as shown in Fig.~\ref{FigSqw}(b). 

\subsubsection{Even and odd distorted coherent states}
\label{gcs21}

\begin{figure}[htb]
\centering
\includegraphics[width=0.25\textwidth]{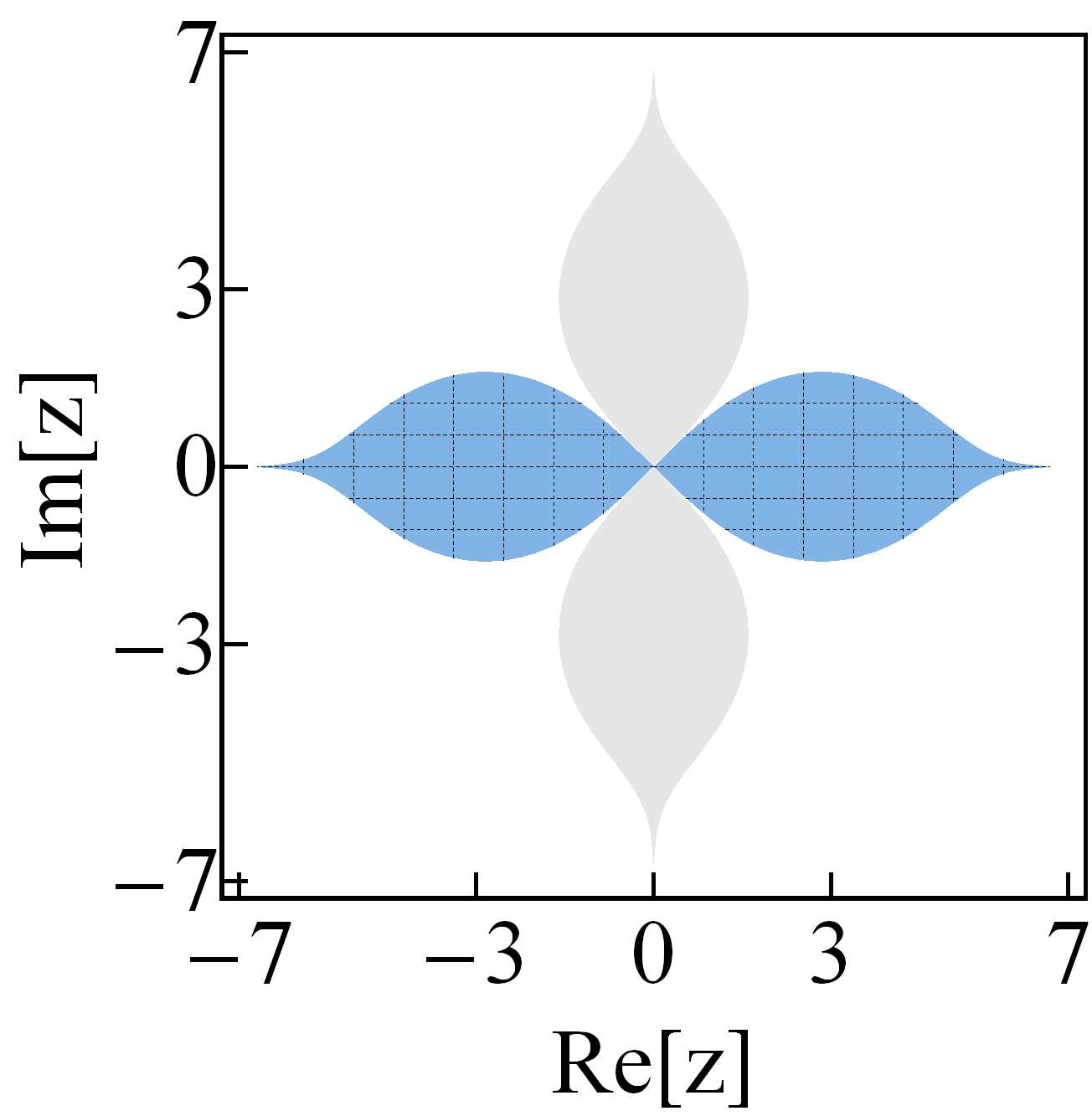}

\caption{\footnotesize 
(Color online) Regions of squeezing in the complex z-plane for the even distorted cat $\vert z_+,w\rangle$ with $w=20$. The grey (mesh, dashed blue) zones correspond to the values of $z \in \mathbb C$ for which $X_w (P_w)$ is squeezed. The white zones indicate the values of $z$ for which both variances are bigger than the average uncertainty. 
}
\label{Quadw}
\end{figure}

For the even and odd distorted cats
\begin{equation}
\vert z_{\pm}, w \rangle = \left[ 1 \pm \frac{ {}_1F_1 (1,w,-\vert z \vert^2)}{{}_1F_1 (1,w,\vert z \vert^2)} \right]^{-1/2} \left( \frac{\vert z,w \rangle \pm \vert -z,w \rangle}{\sqrt 2} \right),
\label{catw}
\end{equation}
the variances of the distorted quadratures (\ref{distquad}) can be expressed in the form given by Eq.~(\ref{c1}) of Appendix~\ref{ApB}, with $A= X_w$, $B=P_w$, and
\begin{equation}
\begin{array}{c}
U_1^{(\pm)} = 2 \left[ \displaystyle\frac{ \left[ \mbox{Re}(z) \right]^2 {}_1F_1(1,w; \vert z \vert^2) \mp  \left[ \mbox{Im}(z) \right]^2 {}_1F_1(1,w; -\vert z \vert^2)}{ {}_1F_1(1,w ; \vert z \vert^2) \pm  {}_1F_1 (1,w ; -\vert z \vert^2) } \right],\\[3ex]
U_2^{(\pm)} = -2 \left[ \displaystyle\frac{ \left[ \mbox{Im}(z) \right]^2 {}_1F_1(1,w; \vert z \vert^2) \mp  \left[ \mbox{Re}(z) \right]^2 {}_1F_1(1,w; -\vert z \vert^2)}{ {}_1F_1(1,w ; \vert z \vert^2) \pm  {}_1F_1 (1,w ; -\vert z \vert^2) } \right].
\end{array}
\end{equation}
In Fig.~\ref{Quadw}, we show the regions of the complex $z$-plane where $X_w$ and $P_w$ are squeezed for the even distorted cat $\vert z_+, w \rangle$ with the indicated value of $w$. Depending on the phase $\phi$, the squeezing is present in zones defined by the interval $0 < \vert z \vert \lesssim 7$ and aligned with either the real or the imaginary axes. Such zones are of maximum width at $\vert z \vert \approx 3.5$ and are slenderer as either $\vert z \vert \rightarrow 0$ or $\vert z \vert \rightarrow \infty$.

\begin{figure}[htb]
\centering
\subfigure[$z=0$]{\includegraphics[width=0.29\textwidth]{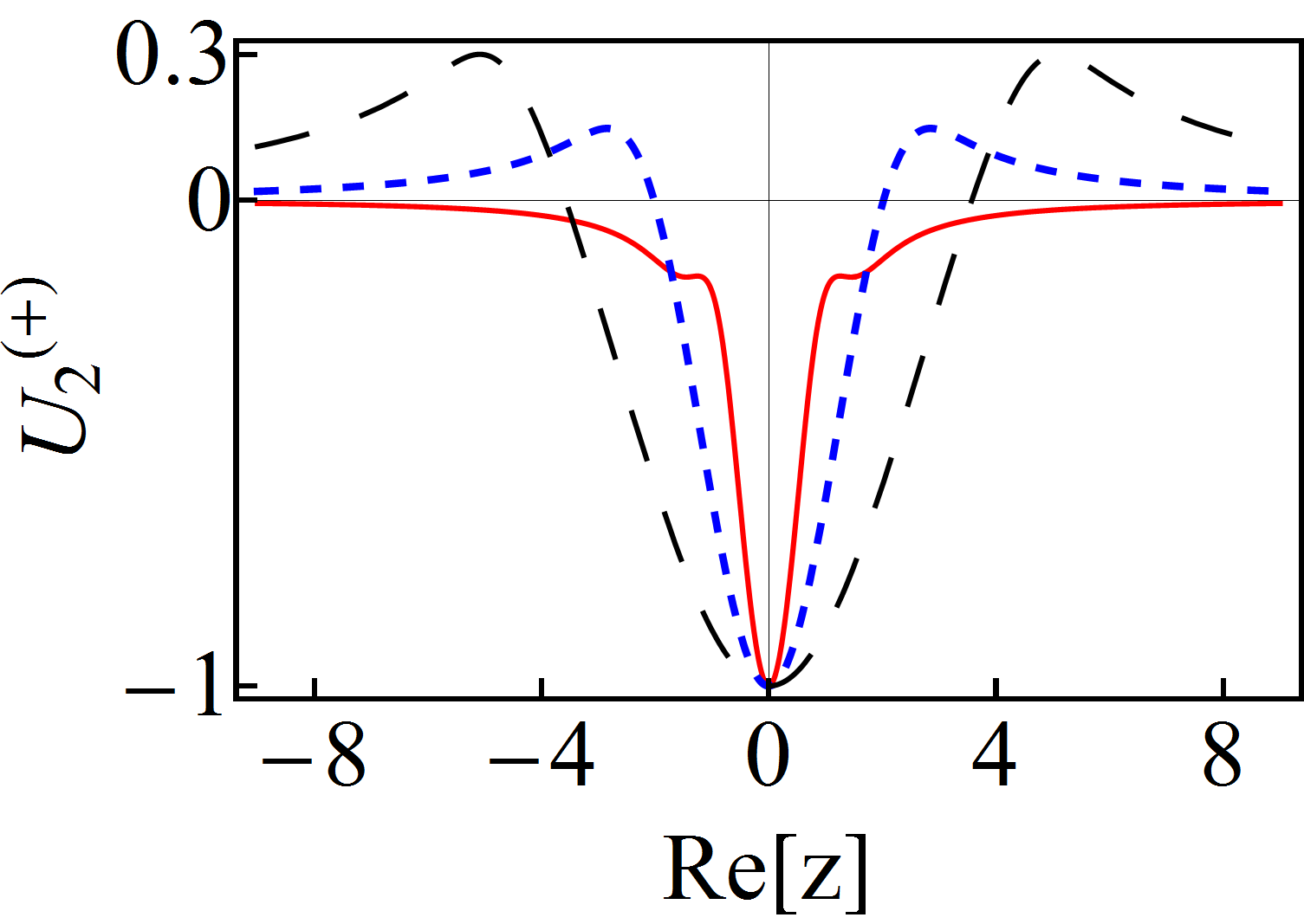}} 
\hskip2ex
\subfigure[$z=0$]{\includegraphics[width=0.3\textwidth]{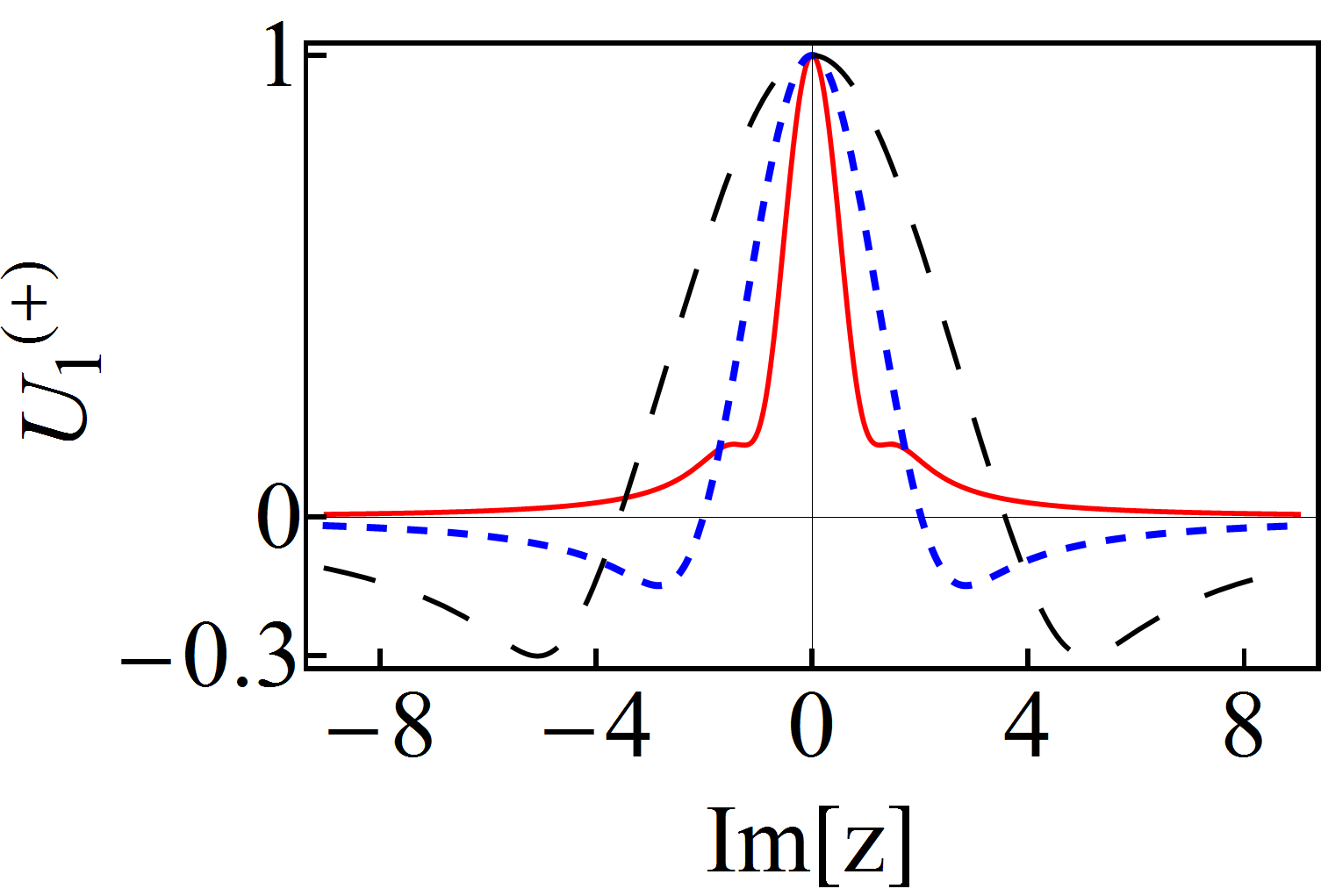}} 
\hskip2ex
\subfigure[$z=0$]{\includegraphics[width=0.27\textwidth]{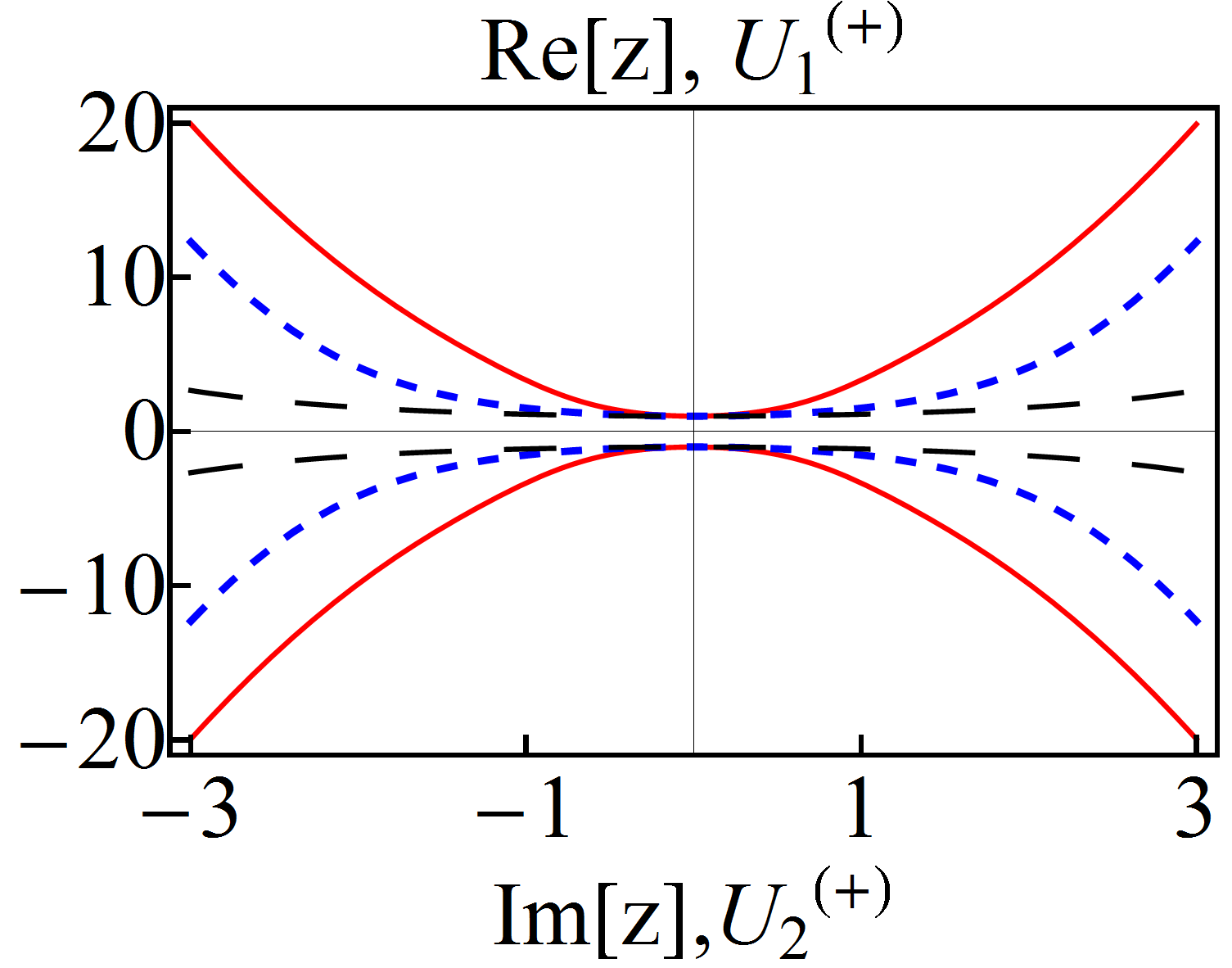}}

\caption{\footnotesize 
(Color online) The parameters $U_1^{(+)}$ and $U_2^{(+)}$ defining the variances of the physical quadratures $\hat x$ and $\hat p$ for the even distorted cat state $\vert z_+^{osc},w \rangle$. In (a) we show the behavior of $U_2^{(+)}$ along the real axis of the complex $z$-plane. In (b) it is shown $U_1^{(+)}$ along the imaginary axis. The graphic (c) shows $U_1^{(+)}$ and $U_2^{(+)}$ along the imaginary and the real axis, respectively. In all the graphics $w=1$ is in solid-red, $w=5$ in dotted-blue, and $w=20$ in dashed-black curves.
}
\label{FigCatw}
\end{figure}

In the oscillator limit (\ref{osc1}), the variances of $\hat x$ and $\hat p$ are calculated using the Eq.~(\ref{c1}) of Appendix~\ref{ApB}, with
\begin{equation}
\begin{aligned}
U_1^{(\pm)}= \frac{\textnormal{Re}(z^2) \left[ f_2 (\vert z \vert^2, w) \pm f_2 (-\vert z \vert^2, w) \right] +\left[ {}_1 F_1\left( 2, \omega;\vert z \vert^{2} \right)  \pm {}_1 F_1\left( 2, \omega; -\vert z \vert^{2} \right)  \right]}{{}_{1}F_{1}(1,\omega;\vert z \vert^{2}) \pm {}_{1}F_{1}(1,\omega;-\vert z \vert^{2})}, \\[2ex]
U_2^{(\pm)}= \frac{\textnormal{Re}(z^2) \left[ f_2 (\vert z \vert^2, w) \pm f_2 (-\vert z \vert^2, w) \right] - \left[ {}_1 F_1\left( 2, \omega;\vert z \vert^{2} \right)  \pm {}_1 F_1\left( 2, \omega; -\vert z \vert^{2} \right)  \right]}{{}_{1}F_{1}(1,\omega;\vert z \vert^{2}) \pm {}_{1}F_{1}(1,\omega;-\vert z \vert^{2})},
\end{aligned}
\end{equation}
and $f_2 (\vert z \vert^2, w)$ given in (\ref{efes}). The quadratures $\hat x$ and $\hat p$ are squeezed in narrow zones of the complex $z$-plane that are defined along the real and imaginary axis, respectively. As in the previous cases, there is no squeezing for small values of $\vert z \vert$. Moreover, the squeezing regions are slenderer as $\vert z \vert \rightarrow  \infty$. To illustrate the phenomenon, in Fig.~\ref{FigCatw} we show the behavior of the $U$-functions that define the variances $(\Delta \hat x)^2$ and $(\Delta \hat p)^2$ of the even distorted cat $\vert z_+^{osc}, w\rangle$. As $U_1^{(+)}$ is positive definite along the real axis, see Fig.~\ref{FigCatw}(c), the quadrature $\hat p$ is squeezed in the intervals of the real axis where the function $U_2^{(+)}$ is non-negative. The latter can be identified in Fig.~\ref{FigCatw}(a). Similarly, from Fig.~\ref{FigCatw}(c) we see that $U_2^{(+)}$ is negative along the imaginary axis. Then, $\hat x$ is squeezed in the intervals of the imaginary axis for which $U_1^{(+)}$ is negative, see Fig.~\ref{FigCatw}(b).

\begin{figure}[htb]
\centering
\subfigure[$\vert z\vert =0$]{\includegraphics[width=0.25\textwidth]{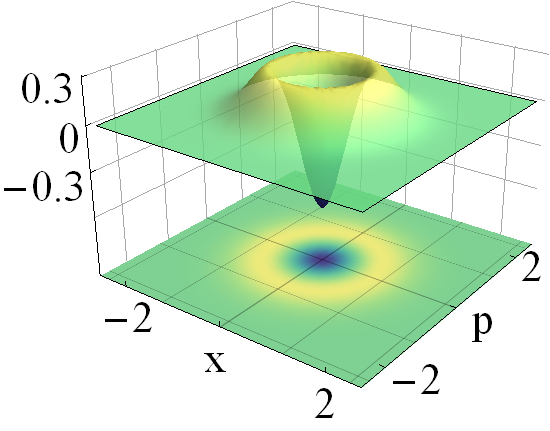}} 
\hspace{1ex}
\subfigure[$\vert z\vert=1.2$]{\includegraphics[width=0.25\textwidth]{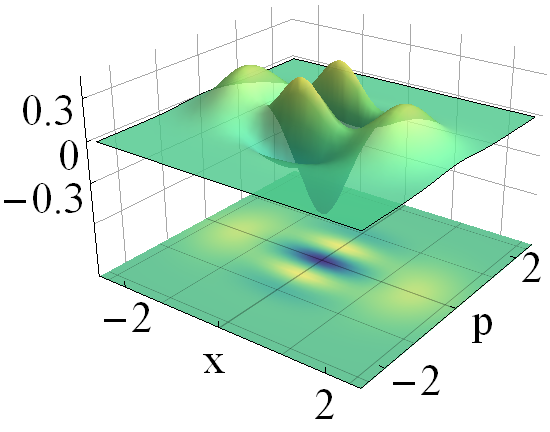}} 
\hspace{1ex}
\subfigure[$\vert z\vert=2$]{\includegraphics[width=0.25\textwidth]{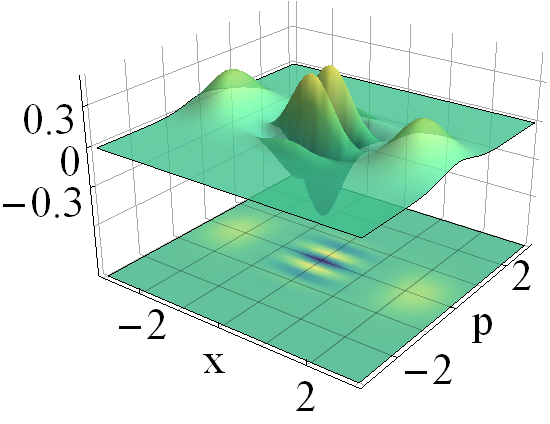}} 

\centering
\subfigure[$\vert z\vert=0$]{\includegraphics[width=0.25\textwidth]{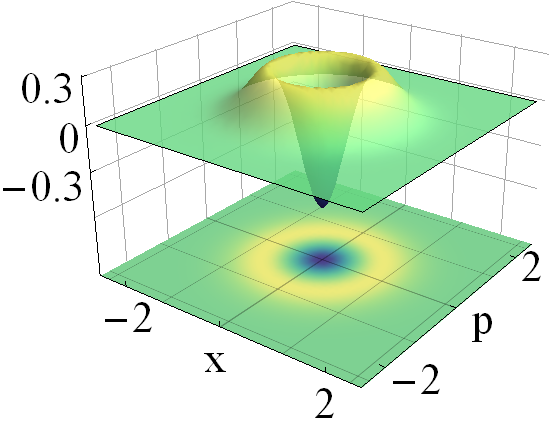}} 
\hspace{1ex}
\subfigure[$\vert z\vert=1.2$]{\includegraphics[width=0.25\textwidth]{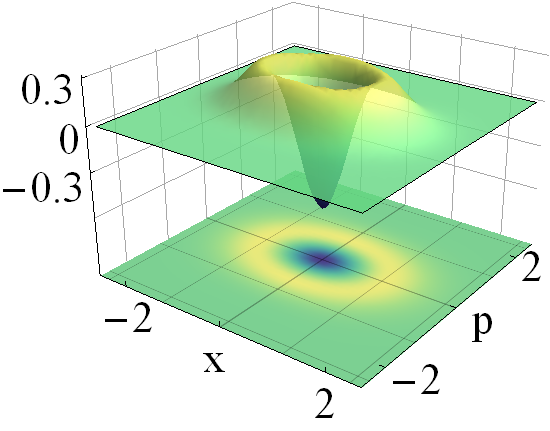}} 
\hspace{1ex}
\subfigure[$\vert z\vert=2$]{\includegraphics[width=0.25\textwidth]{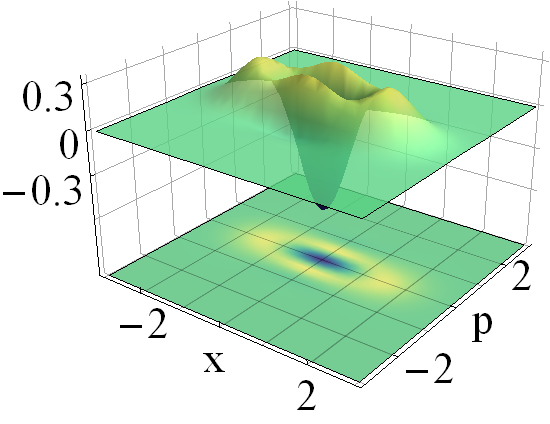}} 

\caption{\footnotesize 
Wigner distribution of the distorted even cat states $\vert z^{osc}_+,w \rangle$ for $w=1$ (upper row), $w=5$ (lower row). Here $z= \vert z \vert e^{i\phi}$, with $\phi =0$, and the indicated values of $\vert z \vert$. 
}
\label{FigWCatw1A}
\end{figure}

\begin{figure}[htb]
\centering
\subfigure[$\vert z\vert=0$]{\includegraphics[width=0.25\textwidth]{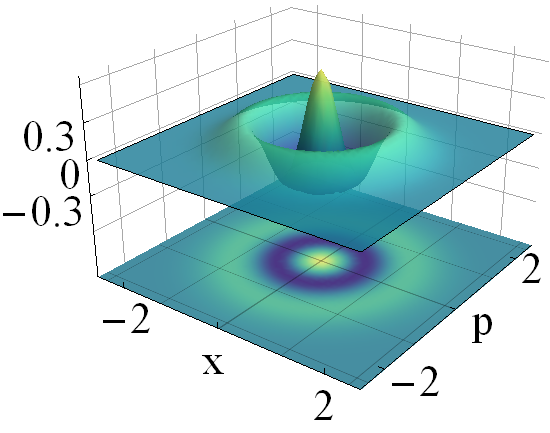}} 
\hspace{1ex}
\subfigure[$\vert z\vert=1.2$]{\includegraphics[width=0.25\textwidth]{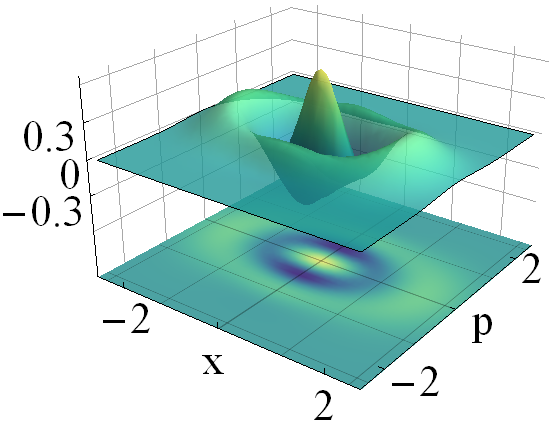}} 
\hspace{1ex}
\subfigure[$\vert z\vert=2$]{\includegraphics[width=0.25\textwidth]{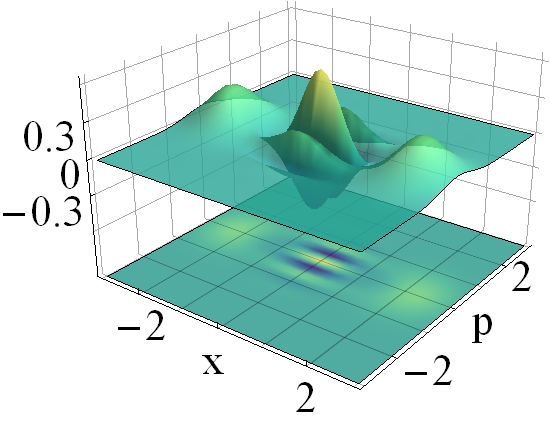}} 

\centering
\subfigure[$\vert z\vert=0$]{\includegraphics[width=0.25\textwidth]{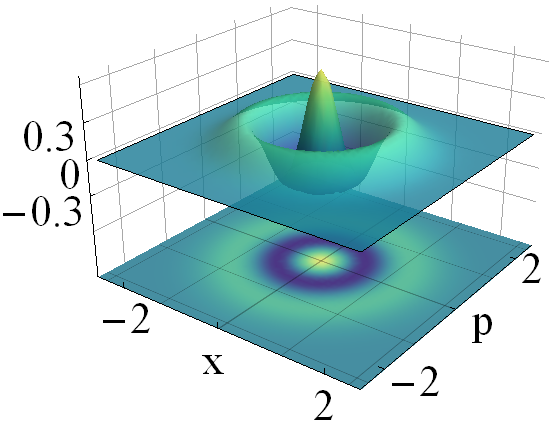}} 
\hspace{1ex}
\subfigure[$\vert z\vert=1.2$]{\includegraphics[width=0.25\textwidth]{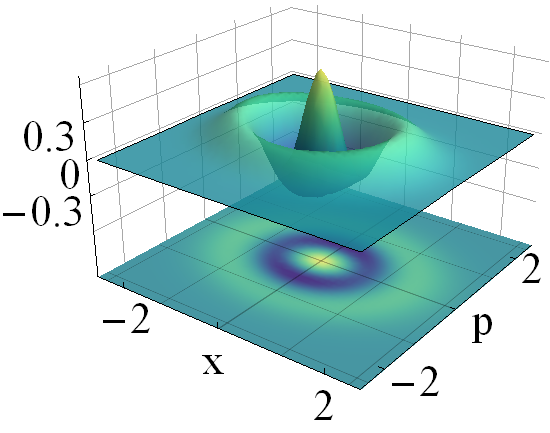}} 
\hspace{1ex}
\subfigure[$\vert z\vert=2$]{\includegraphics[width=0.25\textwidth]{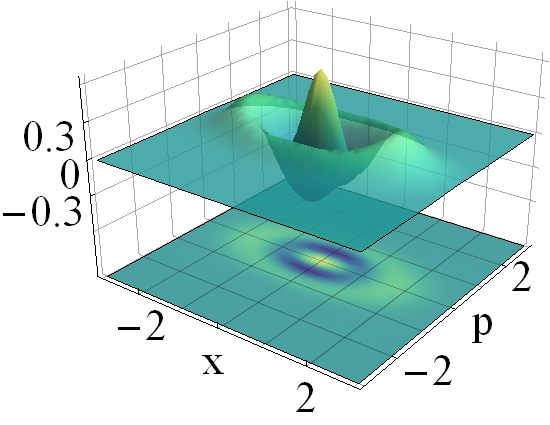}} 

\caption{\footnotesize 
Wigner distribution of the distorted odd cat states $\vert z^{osc}_-,w \rangle$ for $w=1$ (upper row), $w=5$ (lower row). Here $z= \vert z \vert e^{i\phi}$, with $\phi =0$, and the indicated values of $\vert z \vert$. 
}
\label{FigWCatw1B}
\end{figure}

In Figs.~\ref{FigWCatw1A} and \ref{FigWCatw1B} we show the density plot of the Wigner distribution of the even and odd cat states (\ref{catw}), respectively. In both cases we have used $\vert z_{\pm}^{osc}, w\rangle$ for $w=1$, $w=5$, and different values of $z$. The Wigner distribution is negative in different regions of the complex $z$-plane, and it exhibits oscillations as $\vert z \vert$ increases. The distortion parameter $w$ is used to control such oscillations by reducing their number as $w$ increases.

On the other hand, the straightforward calculation shows that neither $X_w$ nor $P_w$ are squeezed for the odd distorted cats $\vert z_-, w \rangle$. The same holds for the physical quadratures $\hat x$ and $\hat p$, evaluated in the oscillator limit state $\vert z^{osc}_-,w \rangle$.

\subsection{Nonclassical displaced coherent states}
\label{gcs3}

Let us analyze the properties of the displaced coherent states $\vert \psi_0 \rangle$ and $\vert z, w \rangle_d$ introduced in (\ref{displaced}). In the oscillator limit, the variances of the physical quadratures $\hat x$ and $\hat p$ are derived from Eq.~(\ref{c1}) of Appendix~\ref{ApB}, with 
\begin{equation}
\begin{aligned}
U_1= \frac{\textnormal{Re}(z^2) h_2 ( \vert z \vert^2, \omega) + {}_2 F_2\left( 2, \omega; 1,1; \vert z \vert^{2} \right) }{{}_1F_1(1,\omega; \vert z \vert^2)} - 2 \left[  \frac{\textnormal{Re}(z) h_1(z, \omega)}{{}_1F_1(1,\omega;\vert z \vert^2) } \right]^2, \\[2ex]
U_2= \frac{\textnormal{Re}(z^2) h_2(\vert z \vert^2, \omega) - {}_2 F_2\left( 2, \omega; 1,1; \vert z \vert^{2} \right) }{{}_1F_1(1,\omega;\vert z \vert^2)}  + 2 \left[ \frac{\textnormal{Im}(z) h_1 (z, \omega)}{{}_1F_1(1,\omega;\vert z \vert^2)} \right]^2,
\end{aligned}
\end{equation} 
and
\begin{equation}
\begin{aligned}
& h_1 (z, \omega)=\sum_{n=0}^{+\infty}\frac{\vert z \vert^{2n}}{n!} \frac{\sqrt{(\omega)_{n}(\omega)_{n+1}(n+2)}}{(n+1)!},\\[2ex] 
& h_2 (\vert z \vert^2, \omega)=\sum_{n=0}^{+\infty} \frac{\vert z \vert^{2n}}{n!} \frac{\sqrt{(\omega)_{n}(\omega)_{n+2}(n+2)(n+3)}}{(n+2)!}.
\end{aligned}
\label{haches}
\end{equation}
The squeezing of $\hat x$ and $\hat p$ for $\vert z^{osc},w \rangle_d$ is quite similar to that obtained for the distorted coherent states of the previous section. 

\begin{figure}[htb]
\centering
\subfigure[$z=0$]{\includegraphics[width=0.3\textwidth]{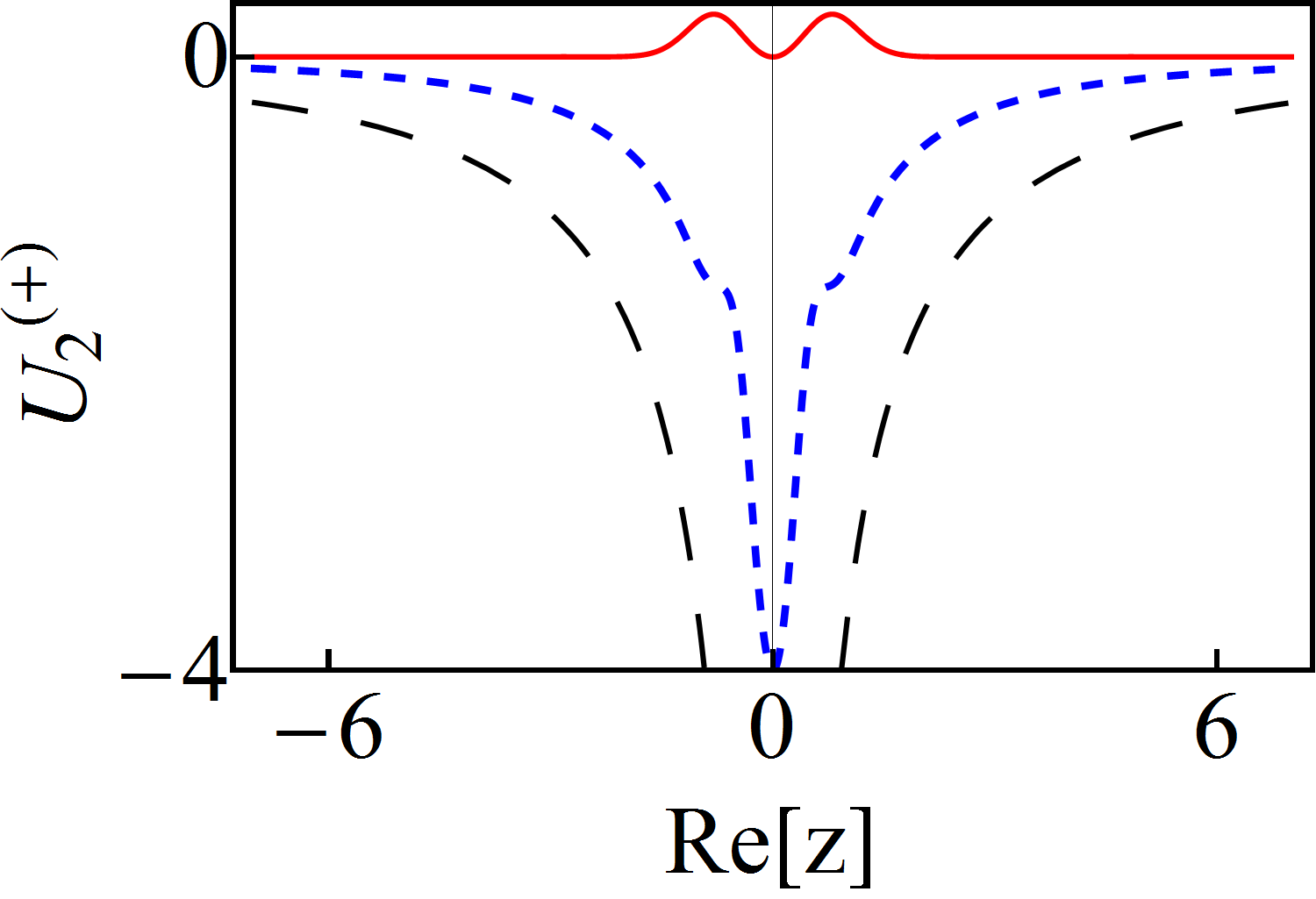}} 
\hskip2ex
\subfigure[$z=0$]{\includegraphics[width=0.28\textwidth]{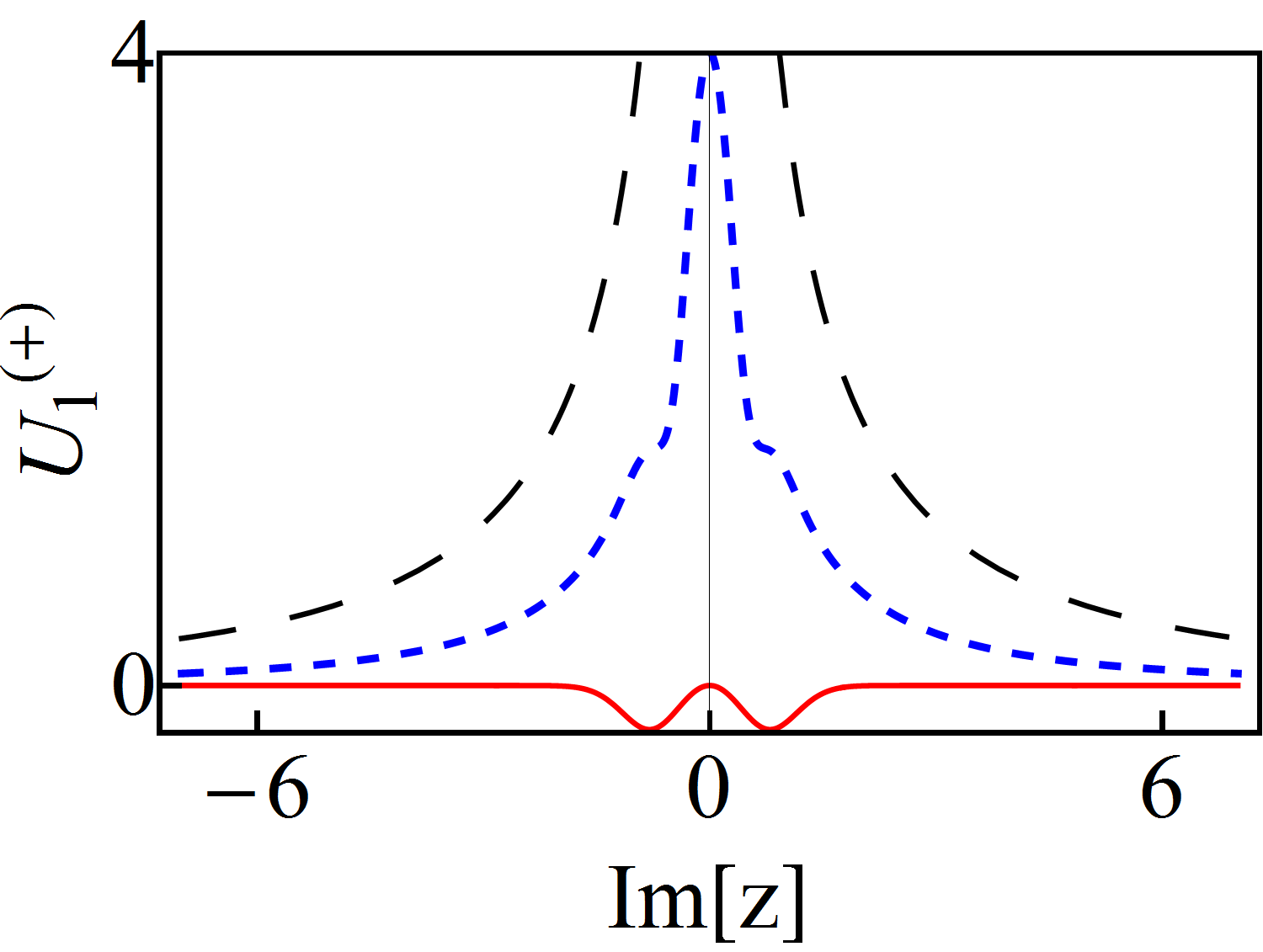}} 
\hskip2ex
\subfigure[$z=0$]{\includegraphics[width=0.27\textwidth]{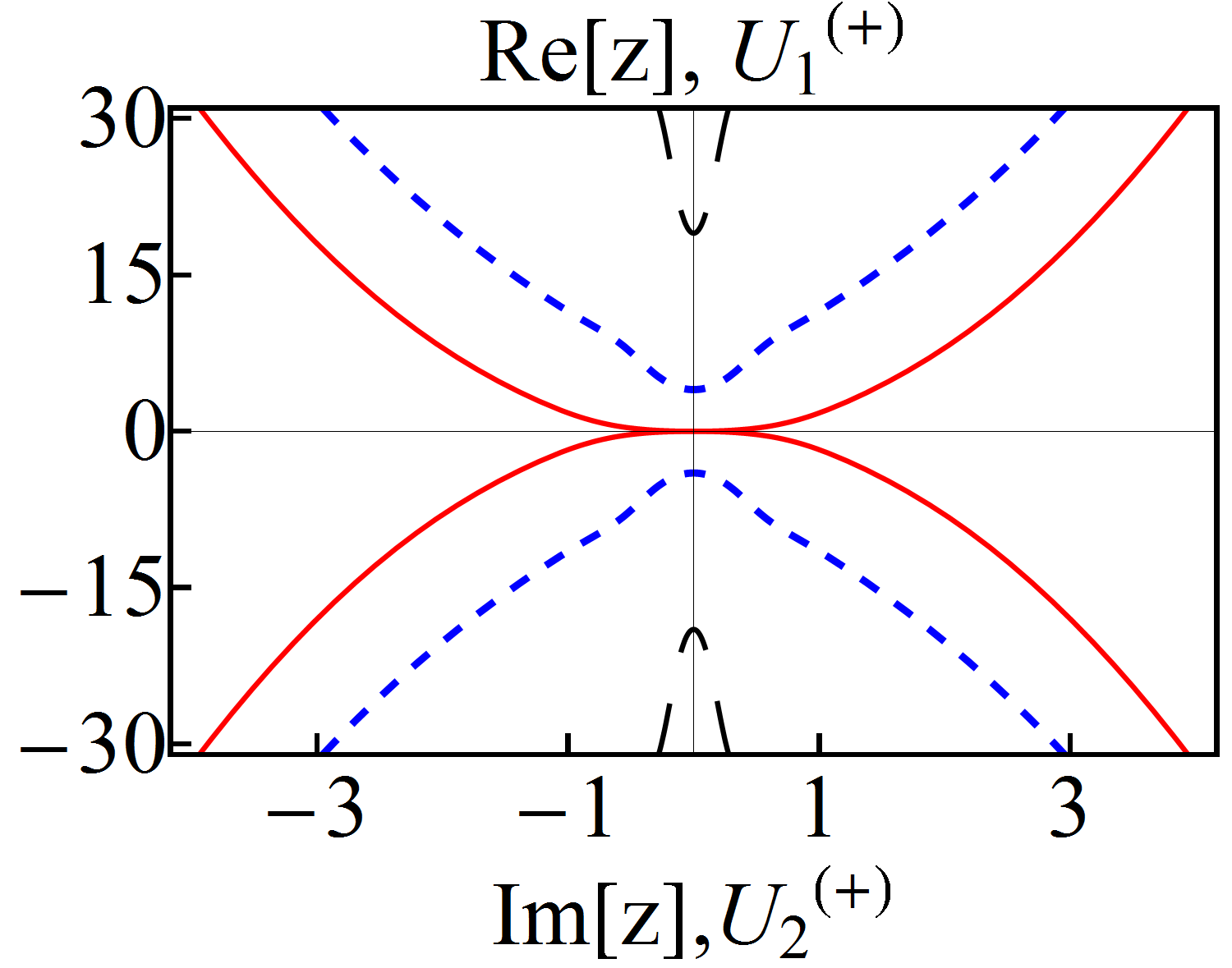}}

\caption{\footnotesize 
(Color online) The parameters $U_1^{(+)}$ and $U_2^{(+)}$ defining the variances of the distorted quadratures $X_w$ and $P_w$ according to Eq.~(\ref{c1}) of Appendix~\ref{ApB} for the even displaced cat state $\vert z_+,w \rangle_d$. In (a) we show the behavior of $U_2^{(+)}$ along the real axis of the complex $z$-plane. In (b) it is shown $U_1^{(+)}$ along the imaginary axis. The graphic (c) shows $U_1^{(+)}$ and $U_2^{(+)}$ along the imaginary and the real axis, respectively. In all the graphics $w=1$ is in solid-red, $w=5$ in dotted-blue, and $w=20$ in dashed-black curves.
}
\label{FigCatD}
\end{figure}

On the other hand, for the even and odd displaced coherent states
\begin{equation}
\vert z_{\pm}, w \rangle_d = \left[ 1 \pm \tfrac{ {}_1F_1 (w,1,-\vert z \vert^2)}{{}_1F_1(w,1,\vert z \vert^2)}
\right]^{-1/2} \left( \vert z,w \rangle_d \pm \vert -z,w \rangle_d \right), 
\label{catd}
\end{equation}
the variances of the distorted quadratures (\ref{distquad}) are calculated from Eq.~(\ref{c1}) of Appendix~\ref{ApB}, with
\begin{equation}
\begin{aligned}
U_1^{(\pm)}= \frac{\textnormal{Re}(z^2) \left[ h_2 (\vert z \vert^2, w) \pm h_2 (-\vert z \vert^2, w) \right] +\left[ {}_2 F_2\left( 2, \omega; 1,1; \vert z \vert^{2} \right)  \pm {}_2 F_2\left( 2, \omega; 1,1; -\vert z \vert^{2} \right)  \right]}{{}_{1}F_{1}(1,\omega;\vert z \vert^{2}) \pm {}_{1}F_{1}(1,\omega;-\vert z \vert^{2})}, \\[2ex]
U_2^{(\pm)}= \frac{\textnormal{Re}(z^2) \left[ h_2 (\vert z \vert^2, w) \pm h_2 (-\vert z \vert^2, w) \right] - \left[ {}_2 F_2\left( 2, \omega; 1,1; \vert z \vert^{2} \right) \pm {}_2 F_2\left( 2, \omega; 1,1; -\vert z \vert^{2} \right)  \right]}{{}_{1}F_{1}(1,\omega;\vert z \vert^{2}) \pm {}_{1}F_{1}(1,\omega;-\vert z \vert^{2})},
\end{aligned}
\nonumber
\end{equation}
and $h_2 (\vert z \vert^2, w)$ given in (\ref{haches}). From Fig.~\ref{FigCatD} we see that, along the real axis, the function $U_1^{(+)}$ is positive definite for any $w$ while $U_2^{(+)}$ is non-negative for $w=1$ only. Similarly, along the imaginary axis, $U_2^{(+)}$ is always negative but only for $w=1$ the function $U_1^{(+)}$ is negative. The conclusion is that only for $w=1$ we obtain the squeezing of either $X_w$ or $P_w$. Remarkably, for this value of the parameter of distortion, the distorted and the displaced coherent states coincide. 

\section{Conclusions}
\label{conclu}

We have studied the nonclassical properties of some pure states of the
non-Hermitian Hamiltonians with real spectrum that are generated by Darboux
(supersymmetric) transformations. Our interest has been focused on
non-Hermitian oscillators since their eigenfunctions form bi-orthogonal bases of states in a straightforward form~\cite{Ros18}. Depending on the parameters that define the Darboux transformation, these oscillators may be either parity-time-symmetric or non-parity-time-symmetric~\cite{Ros15}. The pure states studied in this work are constructed as superpositions of the related bi-orthogonal basis, they include the so-called optimized Binomial and optimized Poisson states~\cite{Zel16,Zel19}, together with the even and odd versions of the generalized coherent states that are associated with the
dynamical algebras of the non-Hermitian oscillators~\cite{Ros18,Zel19}.

One of the main results reported in this paper is to show that the techniques used to study classical like properties for Hermitian systems can be adjusted to investigate the classicality of the states of non-Hermitian systems. The latter is relevant considering that, although the non-Hermitian systems are widely used in contemporary physics~\cite{Moy11,Mos12,Ben07,Fer98,Fak09,Fai87,Sim17}, the classicality of the related quantum states is rarely studied. The main difficulty is that non-Hermiticity of a given operator implies that the orthonormal properties of the related eigenfunctions are not granted a priori. Since most of the approaches implemented to analyze classicality assume orthonormality of states, it is not clear how to translate such methods to the non-Hermitian case in general. Such a difficulty can be overpassed by considering a bi-orthogonal basis for the space of states of the non-Hermitian system, just as we have shown throughout this work. It is expected that our approach can be applied in the study of resonances~\cite{Moy11} (where the non-Hermiticity of the Hamiltonian defining the interaction of a scatterer with a projectile leads to the temporal trapping of the later by the former) and in the propagation of light in materials with complex refractive index~\cite{Mos12} (where the imaginary part of the refractive index may model absorption or amplification of electromagnetic signals).

Concerning the non-Hermitian oscillators we have shown that the corresponding bi-orthogonal bases converge to the Fock basis in the appropriate limit (i.e., after canceling the imaginary part of the potential and properly selecting the parameters), so that the bi-orthogonal superpositions of such states are reduced to pure states of the conventional harmonic oscillator. In particular, we have found that the generalized coherent states of the non-Hermitian oscillators are not minimum uncertainty states for the physical position and momentum quadratures in the oscillator limit. Nevertheless, they exhibit some interesting nonclassical behavior. For instance, depending on the module and complex-phase of the coherence parameter $\alpha$, the physical quadratures can be squeezed for the class of generalized coherent states that we have called ``natural'' (see Section~\ref{seccs}).  The squeezing is clearly exhibited in the corresponding Wigner distribution and revealed by the Mandel parameter for which it is shown that the photon number distribution is sub-Poissonian. The other class of generalized coherent states, called ``distorted'', have similar quadrature squeezing properties. However, the Mandel parameter reveals a different behavior since the photon distribution becomes sub-Poissonian for some values of $\vert\alpha\vert$, but it converges to a Poissonian distribution at the limit $\vert\alpha\vert\rightarrow\infty$. The latter implies a nonclassical-to-classical transition for the distorted coherent states (see \cite{Aga91} for the results on the matter for photon-added coherent states). Such a transition can be also manipulated with the parameter $w$ that defines the related dynamical algebra.

Our method can be extended to construct and study intelligent states associated with the dynamical algebras of the non-Hermitian oscillators. The model involves finite difference equations, which deserve special treatment. Work in this direction is in progress. 


\appendix
\setcounter{section}{0}  
\section{Operator algebras}
\label{ApA}

\renewcommand{\thesection}{A-\arabic{section}}
\setcounter{section}{0}  

\renewcommand{\theequation}{A-\arabic{equation}}
\setcounter{equation}{0}  

The properties of the algebras used throughout this work are summarized below. For the sake of simplicity, hereafter we use the Dirac's notation to represent the states of the system.

$\bullet$ {\bf Quadratic polynomial Heisenberg algebra.} The first pair of ladder operators, ${\cal A}$ and ${\cal A}^+$, together with the Hamiltonian $H_{\lambda}$, satisfy the {\em quadratic polynomial}  (Heisenberg) algebra introduced in Eq.~(\ref{nat2}):
\begin{equation}
{}[{\cal A}, {\cal A}^{+}]= 2 \left(3 H_{\lambda} +1 \right) \left( H_{\lambda} +1 \right), \quad [H_{\lambda}, {\cal A} ]=-2 {\cal A}, \quad [H_{\lambda}, {\cal A}^+ ]= 2 {\cal A}^+. 
\end{equation}
The action of ${\cal A}$ and ${\cal A}^+$ on the eigenvectors of $H_{\lambda}$ is as follows
\begin{equation}
\begin{array}{c}
{\cal A} \vert \psi_{n+1} \rangle = 2n\sqrt{2 (n+1)} \vert \psi_n \rangle, \quad {\cal A}^{+} \vert \psi_{n+1} \rangle = 2(n+1) \sqrt{2 (n+2)} \vert \psi_{n+2} \rangle,\\[2ex]
{\cal A} \vert \psi_0 \rangle = {\cal A}^{+} \vert \psi_0 \rangle =0, \quad n \geq 0.
\end{array}
\label{nat1}
\end{equation}
Thus, ${\cal A}$ annihilates the vectors $\vert \psi_1 \rangle$ and $\vert \psi_0 \rangle$, and ${\cal A}^{+}$ annihilates the vector $\vert \psi_0 \rangle$. In the Hermitian case ($\lambda=0$), the above operators coincide with the generators of the {\em natural} SUSY algebra reported in \cite{Fer07}
\begin{equation}
\left. {\cal A} \right \vert_{\lambda=0} \equiv a^-_{\cal N}, \quad \left. {\cal A}^{+} \right \vert_{\lambda=0} \equiv a^+_{\cal N}.
\label{nat3}
\end{equation}
In the harmonic oscillator limit (\ref{osc1}), they are reduced to the following $f$-oscillator \cite{Man97} ladder operators:
\begin{equation}
{\cal A} \rightarrow \hat a_f =2 \hat N \hat a, \quad {\cal A}^{+} \rightarrow \hat a_f^{\dagger}=2 \hat a^{\dagger} \hat N,
\label{nat4}
\end{equation}
where $\hat N$, $\hat a$, and $\hat a^{\dagger}$ are, respectively, the number, annihilation and creation operators of the (mathematical) harmonic oscillator
\begin{equation}
\hat x = \tfrac12 (\hat a^{\dagger} + \hat a), \quad \hat p = \tfrac{i}2 (\hat a^{\dagger}-\hat a), \quad H_{osc}= \hat p^2 + \hat x^2 = \hat a^{\dagger} \hat a +1 = 2 \hat N+1,
\label{algosc2}
\end{equation}
with
\begin{equation}
[\hat x, \hat p] =i, \qquad \Delta \hat x \Delta \hat p \geq \tfrac12,
\label{algosc3}
\end{equation}
and
\begin{equation}
[\hat a , \hat a^{\dagger} ]=2, \quad [\hat N, \hat a ]= -\hat a, \quad [N, \hat a^{\dagger}] = \hat a^{\dagger}, \quad \hat N =\tfrac12 \hat a^{\dagger} \hat a.
\label{algosc}
\end{equation}
Note that, $\hat a_f$ and $\hat a_f^{\dagger}$ operate on the set $\{ \vert \varphi_n \rangle \}_{n\geq 0}$ quite similar to the form in which ${\cal A}$ and ${\cal A}^{+}$ operate on $\{ \vert \psi_n \rangle \}_{n\geq 0}$. That is, $\hat a_f$ annihilates the vectors $\vert \varphi_1 \rangle$ and $\vert \varphi_0 \rangle$, and $\hat a_f^{\dagger}$ annihilates $\vert \varphi_0 \rangle$. 

One may introduce the quadrature operators corresponding to ${\cal A}$ and ${\cal A}^{+}$:
\begin{equation}
X_{\cal N}= \frac12 \left( {\cal A}^{+} + {\cal A}\right), \quad P_{\cal N}= \frac{i}2 \left( {\cal A}^{+} - {\cal A} \right),
\label{quadnat}
\end{equation}
which we call the `natural quadratures'. They satisfy
\begin{equation}
\begin{array}{c}
[ X_{\cal N}, P_{\cal N} ]  =\tfrac{i}2 [{\cal A}, {\cal A}^{+}] =i \left(3 H_{\lambda} + 1 \right) \left( H_{\lambda} +1 \right), \\[2ex]
\Delta X_{\cal N} \Delta P_{\cal N} \geq \tfrac12 \left\vert \langle (3H_{\lambda} +1)(H_{\lambda} +1) \rangle \right\vert.
\end{array}
\label{quadnat2}
\end{equation}

$\bullet$ {\bf Distorted Heisenberg algebra.} The second pair of ladder operators, denoted by ${\cal C}_w$ and ${\cal C}_w^{+}$, together with the Hamiltonian $H_{\lambda}$, and an additional operator $I_w$, satisfy the {\em distorted} (Heisenberg) algebra introduced in Eq.~(\ref{dist1}):
\[
[{\cal C}_w, {\cal C}_w^{+} ] = I_w, \quad [H_{\lambda}, {\cal C}_w ]= -2 {\cal C}_w, \quad [H_{\lambda}, {\cal C}_w^+ ]= 2 {\cal C}_w^+.
\]
The action of ${\cal C}_w$, ${\cal C}_w^{+}$ and $I_w$ on the eigenvectors of $H_{\lambda}$ is as follows
\begin{eqnarray}
{\cal C}_w \vert \psi_n \rangle = (1-\delta_{n,0}-\delta_{n,1}) \sqrt{n-2+w} \vert \psi_{n-1} \rangle,
\label{dist2a}\\[2ex]  
{\cal C}_w^{+} \vert \psi_n \rangle = (1-\delta_{n,0}) \sqrt{n-1+w} \vert \psi_{n+1} \rangle,
\label{dist2b} \\[2ex]
I_w \vert \psi_n \rangle = [1-\delta_{n,0} + \delta_{n,1} (w-1) ] \vert \psi_n \rangle,
\label{dist2c}
\end{eqnarray}
where $n\geq 0$ and $w$ is a non-negative parameter that defines the `distortion' of the oscillator algebra (\ref{algosc}). In the Hermitian case, these operators coincide with the generators of the {\em distorted} SUSY algebra reported in \cite{Fer95,Ros96},
\begin{equation}
\left. {\cal C}_w \right\vert_{\lambda=0} =C_w, \quad \left. {\cal C}_w^{+} \right\vert_{\lambda=0} =C_w^{\dagger}.
\label{dist3}
\end{equation}
In the harmonic oscillator limit one has the $f$-oscillator ladder operators
\begin{equation}
{\cal C}_w \rightarrow \hat c_w= \frac{1}{2N}\sqrt{ \frac{N+w-1}{N+1}} \, \hat a_{\cal N}, \quad {\cal C}_w^{+} \rightarrow \hat c_w^{\dagger}= \frac{1}{2(N-1)} \sqrt{ \frac{N+w-2}{N}} \, \hat a_{\cal N}^{\dagger}.
\label{dist4}
\end{equation}
As in the previous case, $\hat c_w$ annihilates the vectors $\vert \varphi_1 \rangle$ and $\vert \varphi_0 \rangle$ while $\hat c_w^{\dagger}$ annihilates $\vert \varphi_0 \rangle$. The corresponding quadrature operators are given by
\begin{equation}
X_w = \frac12 \left( {\cal C}_w^{+} + {\cal C}_w \right), \quad  P_w = \frac{i}2 \left( {\cal C}_w^{+} - {\cal C}_w \right),
\label{distquad}
\end{equation}
which will be called `distorted quadratures' and they satisfy 
\begin{equation}
[ X_w,  P_w] =\tfrac{i}2 [{\cal C}_w, {\cal C}_w^{+}] =\tfrac{i}2 I_w, \qquad \Delta X _w \Delta  P_w \geq \tfrac14 \vert \langle I_w \rangle \vert.
\label{distquad2}
\end{equation}

\appendix
\setcounter{section}{1}  
\section{Nonclassicality criteria}
\label{ApB}

\renewcommand{\thesection}{B-\arabic{section}}

\renewcommand{\theequation}{B-\arabic{equation}}

The criteria used to analyze the nonclassicality of the superpositions throughout this work are the following.

$\bullet$ {\bf Squeezing.} For any two operators $A$ and $B$ with commutator $[A,B]=i C$, the variances can be expressed as
\begin{equation}
(\Delta A)^2= \langle A^2\rangle-\langle A \rangle^2=
\tfrac12 \vert \langle C \rangle \vert + U_1, \quad 
(\Delta B)^2=  \langle B^2\rangle-\langle B \rangle^2=
\tfrac12 \vert \langle C \rangle \vert - U_2.
\label{c1}
\end{equation}
If $U_1$ and $U_2$ are both equal to zero then the root-mean-square deviations become equal $\Delta A = \Delta B= \sqrt{\tfrac12 \vert \langle C \rangle \vert}$, and the uncertainty relationship between $A$ and $B$ is minimized, $\Delta A  \Delta B=\tfrac12 \vert \langle C \rangle \vert$. If $U_1 \neq U_2 \neq0$ we have two different cases (i) $U_1$ and $U_2$ are both positive, then $\Delta A > \Delta B$ and we say that $B$ is squeezed (ii) $U_1$ and $U_2$ are both negative, then $\Delta A < \Delta B$ and we say that $A$ is squeezed. This criterion is used along the paper to analyze the inequalities (\ref{quadnat2}) and (\ref{distquad2}) in their respective state spaces.

On the other hand, it is well known that the Mandel parameter \cite{Man79}
\begin{equation}
Q= \frac{(\Delta N)^2}{\langle N \rangle} -1,
\label{mandel}
\end{equation}
dictates a sub-Poissonian, Poissonian and super-Poissonian photon number distribution for $-1\leq Q<0$, $Q=0$ and $Q>0$, respectively. Nonclassicality corresponds to the sub-Poissonian distributions, which is associated with the squeezing of the photon number $0\leq (\Delta N)^2 < \langle N \rangle$.

$\bullet$ {\bf Beam-splitter technique.} The action of a beam splitter on a given state $\vert \mbox{in} \rangle = \vert \Phi_1 \rangle \otimes \vert \Phi_2 \rangle$ is that it produces non-separable outputs $\vert \mbox{out} \rangle$ in general \cite{Kim02}. If $\vert \mbox{out} \rangle$ is entangled, then the signal $\vert \Phi_1\rangle$ is nonclassical, even if the ancilla $\vert \Phi_2 \rangle$ is a classical state. The latter criterion is used with $\vert \Phi_2\rangle = \vert \varphi_0 \rangle$, which is classical, and $\vert \Phi_1\rangle$ being any of the bi-orthogonal superpositions at the oscillator limit (\ref{osc1}). Thus, we may write 
\begin{equation}
\vert \mbox{in}^{osc} \rangle = \sum_{k=0}^K c_k \vert \varphi_{k+r} \rangle \otimes \vert \varphi_0 \rangle,
\end{equation}
where the super-label ``$osc$'' means that the oscillator limit (\ref{osc1}) has been applied. The action of the beam-splitter is represented by the unitary operator \cite{Kim02}
\begin{equation}
BS=\exp \left[ \frac{\theta}2 \left( a_1^{\dagger} \otimes a_2 e^{i\phi} - a_1 \otimes a_2^{\dagger} e^{-i\phi}
\right)
\right],
\end{equation}
where $a_k$ and $a_k^{\dagger}$ are the ladder operators acting on the  input states $\vert \Phi_k^{osc} \rangle$, with $k=1,2$. Up to a phase, the reflection and transmission coefficients of the beam-splitter are respectively given by $R=\sin \tfrac{\theta}2$ and $T=\cos \tfrac{\theta}2$, with $\theta \in [0,\pi)$. Using $\vert \textnormal{in}^{osc} \rangle$, the straightforward calculation shows that the output state is of the form
\begin{equation}
\vert \mbox{out}^{osc} \rangle = \left( \sum_{k=0}^{K+r} \sum_{p=0}^{\,\,K+r-k} -\sum_{k=0}^{r-1} \sum_{p=0}^{\,\,r-1-k}
\right) \Gamma_{k,p} \vert \varphi_k \rangle \otimes \vert \varphi_p \rangle,
\end{equation}
where
\begin{equation}
\Gamma_{k,p}= e^{-i(\phi-\pi)p} R^p T^k \sqrt{\frac{(k+p)!}{k!p!} }\,  c_{k+p-r}.
\end{equation}
The {\em purity} (linear entropy) $S_L(\rho_1)=1- \mbox{Tr}(\rho_1^2)$ of the signal state $\rho_1 = \mbox{Tr}_2 (\vert \mbox{out}^{osc} \rangle \langle \mbox{out}^{osc} \vert)$ is given by 
\begin{equation}
\begin{array}{rl}
S_L(\rho_1)=1- \mbox{Tr} (\rho_1^2)  & =\displaystyle\sum_{n,m=0}^K \vert F_{n+r,m+r} \vert^2 + \sum_{n,m=0}^{r-1} \vert F_{n,m} + G_{n,m} -H_{n,m} -H_{m,n}^* \vert^2\\[2.5ex]
& \qquad + 2 \displaystyle\sum_{n=0}^{r-1} \sum_{m=0}^K \vert F_{n,m+r}- H_{m+r,n}^* \vert^2,
\end{array}
\label{rho1}
\end{equation}
with
\begin{equation}
\begin{array}{c}
F_{n,m} = \displaystyle\sum_{p=0}^{K+r-\text{Max}\{n,m\} } \Gamma_{n,p} \Gamma_{m,p}^*, \qquad 
G_{n,m} = \displaystyle\sum_{p=0}^{r-1-\text{Max}\{n,m\} } \Gamma_{n,p} \Gamma_{m,p}^*,\\[2.5ex]
H_{n,m} =\displaystyle \sum_{p=0}^{\text{Max}\{K+r-n,r-1-m\} } \Gamma_{n,p} \Gamma_{m,p}^*.
\end{array}
\label{rho2}
\end{equation}
Classical states satisfy the separability condition $S_L=0$ while the maximal entanglement is obtained for $S_L=1$. Then, the nonclassicality is associated with $0< S_L \leq 1$. 

The results for $\vert \phi^{(\gamma)} \rangle$ are obtained by making $r=0$, at the limit $K \rightarrow +\infty$. In such case,  the last two additive terms in (\ref{rho1}) are equal to zero, so that $S_L(\rho_1)$ acquires the form of the linear entropy studied in e.g. \cite{San15,San16}.

$\bullet$ {\bf Wigner function.}  In the basis of the Glauber states \cite{Gla07}:
\begin{equation}
\vert \alpha_G \rangle = e^{-\vert \alpha \vert^2/2} \sum_{k=0}^{+\infty} \frac{\alpha^k}{\sqrt{k!} } \vert \varphi_k \rangle, \quad \alpha \in \mathbb C,
\label{glauber}
\end{equation}
the Wigner function \cite{Wig32} of the state $\rho_{\phi} =\vert \phi \rangle \langle \phi \vert$, with $\vert \phi \rangle$ given either by (\ref{finite}) or (\ref{super2}), is expressed as
\begin{equation}
W(\alpha,\phi) = \tfrac2{\pi} e^{2\vert \alpha \vert^2}  \int d^2 \beta \, \langle -\beta \vert \rho_{\phi}  \vert \beta \rangle e^{2(\beta^* \alpha - \beta \alpha^*)}, \quad \beta \in \mathbb C.
\label{wigner}
\end{equation}
If the Wigner function $W(\alpha, \phi)$ is negative in at least a definite region of the phase-space, then the state $\rho_{\phi}$ is nonclassical.

\section*{Acknowledgments}

The financial support from CONACyT (Mexico), grant number A1-S-24569, MINECO (Spain), project MTM2014- 57129-C2-1-P, and Junta de Castilla y Le\'on (Spain), project VA057U16, is acknowledged. K.~Zelaya is supported by a postdoctoral fellowship from the Mathematical Physics Laboratory, Centre de Recherches Math\'ematiques. S.~Dey acknowledges the support of research grant (DST/INSPIRE/04/2016/001391) from the DST, Govt. of India. V.~Hussin acknowledges the support of research grants from NSERC of Canada.


\end{document}